\documentclass{article}
 
 \usepackage[a4paper,left=2cm,right=2cm,bottom=2cm,top=2cm]{geometry}

 \usepackage{setspace} 
\usepackage{moreverb,url}

\usepackage{authblk}

\usepackage{graphicx}

\usepackage{amsmath}
\usepackage{amssymb}
\usepackage{amsthm}
\usepackage{bm}
\usepackage{bbm}
\usepackage{float}

\usepackage{listings}

\def\bSigma{{\bm\Sigma}}

\def\btheta{{\bm\theta}}
\def\bTheta{{\bm\Theta}}
\def\bT{{\bm{T}}}
\def\bV{{\bm{V}}}
\def\bZ{{\bm{Z}}}

\def\DD{{\mathcal{D}}}

\def\cov{\mathrm{cov}}
 
\def\asim{\stackrel{appr.}{\sim}}

\begin{document}

\title{Simultaneous inference procedures for the comparison of multiple characteristics of two survival functions}

\author[1,*]{Robin Ristl}
\author[2]{Heiko G\"otte}
\author[2]{Armin Sch\"uler}
\author[1]{Martin Posch}
\author[1]{Franz K\"onig} 

\affil[1]{Medical University of Vienna, Center for Medical Data Science, Vienna, Austria}
\affil[2]{Merck KGaA, Darmstadt, Germany}
\affil[*]{Correspondence: robin.ristl@meduniwien.ac.at}

\maketitle
\begin{abstract}
\noindent Survival time is the primary endpoint of many randomized controlled trials, and a treatment effect is typically quantified by the hazard ratio under the assumption of proportional hazards. Awareness is increasing that in many settings this assumption is a-priori violated, e.g. due to delayed onset of drug effect. In these cases, interpretation of the hazard ratio estimate is ambiguous and statistical inference for alternative parameters to quantify a treatment effect is warranted. We consider differences or ratios of milestone survival probabilities or quantiles, differences in restricted mean survival times and an average hazard ratio to be of interest. Typically, more than one such parameter needs to be reported to assess possible treatment benefits, and in confirmatory trials the according inferential procedures need to be adjusted for multiplicity. By using the counting process representation of the mentioned parameters, we show that their estimates are asymptotically multivariate normal and we propose according parametric multiple testing procedures and simultaneous confidence intervals. Also, the logrank test may be included in the framework. Finite sample type I error rate and power are studied by simulation. The methods are illustrated with an example from oncology. A software implementation is provided in the R package nph.\\

\end{abstract}

\section{Introduction}
The aim of survival analysis in a randomized clinical trial typically is to show a benefit of an experimental treatment over a control treatment. Under the frequent model assumption of proportional hazards, the relative treatment benefit can be quantified with a single parameter, the hazard ratio, which describes the shift of the survival function under treatment compared to the survival function under control at every time-point.

When aiming to establish superiority of a new treatment over control in terms of survival in a randomized controlled clinical trial, the null hypothesis of equal survival functions is typically tested by a logrank test, or an equivalent test based on the Cox model, and a hazard ratio estimate is reported to quantify the treatment effect. Under the assumption of proportional hazards, this approach is efficient in terms of power and 
allows for unambiguous conclusions on superiority, because one single parameter, the hazard ratio, is sufficient to describe the shift between survival functions at every time-point.

Recently, the assumption of proportional hazards has been found to be inadequate or at least questionable in relevant clinical settings \cite{finke2007lessons,workshopFDA2018,freidlin2019methods,lin2020alternative,ristl2021delayed}. In particular, a delayed onset of treatment effect in immuno-oncology drugs has been discussed as important source of non-proportional hazards and, moreover, as a setting in which the traditional logrank test looses power and the effect quantification using the hazard ratio as a single summary measure may be flawed. Other sources of non-proportional hazards that have been discussed in the literature are, e.g., a heterogeneous patient population with population subgroups that respond differently to treatment, such as long-term survivors, modified efficacy after disease progression, or the need for rescue medication or treatment switching 
\cite{ristl2021delayed,shen2021nonproportional,ananthakrishnan2021critical}.

Several testing procedures have been proposed to remedy the potential loss of power of the logrank test in these settings. In particular, the use of weighted logrank tests has been proposed, putting more weight on event times with a more pronounced anticipated effect \cite{fleming1991counting,magirr2019modestly}. If the pattern of non-proportional hazards is not well known beforehand, a maximum combination (MaxCombo) test of differently weighted logrank tests has been shown to be a robustly powerful method \cite{tarone1981distribution,lee1996some,karrison2016versatile,ristl2021delayed,royston2020simulation,ghosh2022robust}. However, in absence of further assumptions, a significant result for these tests, at first, only implies that there exists a time point for which the hazard function under treatment is less than the hazard function under control. Whether this result translates into a relevant (or in fact into any \cite{magirr2019modestly}) benefit in terms of the survival functions needs to be assessed by estimates for parameters that appropriately quantify the differences between survival functions \cite{shen2021nonproportional,posch2022testing}. 

The interpretation of the parameter estimate of the Cox model can be challenging under non-proportional hazards. In particular, the limiting value of the Cox model hazard ratio estimate will depend not only on the true hazard functions but also on study design parameters such as recruitment rate, study duration and censoring distribution that affect the timing of observed events. Alternative effect measures, proposed to be used under non-proportional hazards, include the difference in restricted mean survival times \cite{royston2011use,royston2013restricted}, average hazard ratios defined via predefined weighting functions \cite{kalbfleisch1981estimation,schemper2009estimation,rauch2018average}, differences x-year milestone survival probabilities or differences in quantiles of the survival distribution. Several authors have argued that a single such parameter may not be sufficient and differences in survival curves should instead be assessed by a set of summary statistics \cite{lin2020alternative,shen2021nonproportional,roychoudhury2021robust}.

In this paper, we propose a simultaneous inference framework for a set of multiple parameters, which may include differences in survival probabilities, differences in log survival probabilities, differences in complementary log log (cloglog) transformed survival probabilities, differences in quantiles of the survival functions, differences in log transformed quantiles, an average hazard ratio and the difference in restricted mean survival times. The logrank test, albeit being a non-parametric test, may be included, too. For completeness, the Cox model hazard ratio may be included under the assumption of proportional hazards.

We present multiple testing procedures and simultaneous confidence intervals for pre-specified sets of parameters, such that the family wise type I error rate is controlled and multiple parameter estimates can be interpreted simultaneously in confirmatory manner. The multiple testing adjustment is based on the counting process representation of survival function estimates, by which we show that the considered estimates are asymptotically multivariate normal and by which we derive an estimate of their asymptotic covariance matrix.

The paper is structured as follows. In Section 2, we propose a  general framework for the multivariate normal approximation of multiple survival parameters and show the application to particular estimates. 
In Section 3, we describe multiple testing procedures and simultaneous confidence intervals, In Section 4, we perform a simulation study to assess the operating characteristics of the proposed methods in terms of type I error rate and power under different scenarios. In Section 5, we illustrate the proposed methods in a worked example based on survival curves 
reported by Burtness et al. \cite{burtness2019pembrolizumab} (KEYNOTE-048)  for the comparison of pembrolizumab versus cetuximab in treating recurrent or metastatic squamous squamous cell carcinoma. A software implementation of all proposed methods is implemented in the R package nph \cite{ristl2021delayed} (see Section 6). We conclude with a discussion.

\section{Multivariate normal approximation for multiple survival parameter estimates}
In this section we present a general framework for a multivariate normal approximation and the estimation of the covariance matrix for multiple parameters derived from survival functions.
We subsequently apply the framework to different parameter estimates that are commonly used to quantify the difference between two survival curves.

\subsection{General framework based on martingale representation}
\label{sec.asym}
We consider a control and a treatment group, indexed $i=0,1$, including $n_i$ independent subjects with observations on possibly censored survival times. Denote with $\DD_i$ the set of observed event times in group $i$. We assume that the censoring times are stochastically independent of the true event times.

For further notation purposes, let $S_i(t)=P_i(T_i>t)$ be the survival function in group $i$, with $T_i$ a random event time from the respective group. The corresponding hazard function is $\lambda_i(t)=\frac{dS_i(t)}{dt}\frac{1}{S_i(t)}$, and the cumulative hazard function is $\Lambda_i(t) = \int_0^t \lambda_i(s)ds = -\log(S_i(t))$. Further, for a given data sample, let $N_i(t)$ be the number of observed events in the time interval $[0,t]$ and $Y_i(t)$ the number at risk at time $t$ in group $i$. We denote by $\hat\Lambda_i(t)=\sum_{s\in \DD_i,s\leq t} dN_i(s)/Y_i(s) $ the Nelson-Aalen estimator for the cumulative hazard, by $\hat S_i(t)=\exp\{-\hat\Lambda_i(t)\}$ the Nelson-Aalen estimator for $S_i(t)$ and by $\hat S_i^{-}(t)=\exp\{-\sum_{s\in \DD_i,s < t} dN_i(s)/Y_i(s)\}$ the left continuous version of $\hat S_i(t)$. 
Define $\hat{q}_i(\gamma)=\min\{t:\hat{S}_i(t)\leq 1-\gamma\}$
as an estimate of the $\gamma$-quantile of the survival distribution in group $i$.

Further we denote by $M_i(t)=N_i(t)-\int_0^t Y_i(s)d\Lambda_i(s)$ the difference between observed and expected events up to time $t$. $M_i(t)$ is a martingale process that is key to the counting process representation of estimates in survival analysis \cite{aalen2010history,fleming1991counting}. 

As a general framework, we aim to quantify the difference in the survival functions under control and treatment by a set of parameters $\theta_1,\hdots,\theta_m$. We assume that
the true parameters are of the form $\theta_k = \theta_{k,1} - \theta_{k,0}$, where $\theta_{k,i}$ is a function of the true survival function and possibly of the true censoring function in group $i$. 

We further assume that the difference between the corresponding estimates $\hat\theta_{k,i}$ and the true parameter values $\theta_{k,i}$ can be approximated up to an asymptotically negligible residual term by a stochastic integral such that 
\begin{equation}\label{eq:martingale_repr}
\hat\theta_{k,i} -\theta_{k,i}  = a_{k,i}\int_0^{t_k} H_{k,i}(s)\frac{1}{Y_i(s)} dM_i(s) + o_p(1/\sqrt{n_i})
\end{equation}
where the $a_{k,i}$ are constant parameters for which consistent estimates $\hat{a}_{k,i}$ exist and the $H_{k,i}$ are predictable processes with respect to the martingale process $M_i$ for which consistent estimates $\hat{H}_{k,i}$ exist. The interval $[0,t_k]$ is the time interval within which data is used to calculate $\hat\theta_{k,i}$.

We also assume that
$\hat\theta_{k,0}$ and $\hat\theta_{k',1}$ are asymptotically uncorrelated for all $k=1,\hdots,m$ and $k'=1,\hdots,m$.

Finally we assume the asymptotic stability condition holds \cite{mckeague1988stochastic}: There exists a function $\rho_i$ with values in $(0,1)$ such that for $n_i \rightarrow \infty$, $\sup_{0<s\leq \max(t_1,\hdots,t_m)} |\frac{Y_i(s)}{n_i}-\rho_i(s)| \rightarrow 0$ a.s.

Then, by the multivariate martingale central limit theorem, the vector

$(\sqrt{n_i}(\hat\theta_{1,i}-\theta_{1,i}),\hdots,\sqrt{n_i}(\hat\theta_{m,i}-
\theta_{m,i}))$ asymptotically follows a multivariate normal distribution with mean zero and variances and covariances given by \cite{fleming1991counting} \[\cov\left(\sqrt{n_i}(\hat\theta_{k,i}-\theta_{k,i}),\sqrt{n_i}(\hat\theta_{k',i}-
\theta_{k',i})\right)=a_{k,i}a_{k',i}\int_0^{t_{k}\wedge t_{k'}} H_{k,i}(s)H_{k',i}(s)\frac{1}{\rho_i(s)} d\Lambda(s)) .\]

A consistent estimator $\hat\Sigma_i$ for the covariance matrix of  $(\hat\theta_{1,i},\hdots,\hat\theta_{m,i})$ is obtained by replacing $a_{k,i}$ by $\hat{a}_{k,i}$, $H_{k,i}$ by $\hat{H}_{k,i}$, $\rho_i(s)$ by $Y_i(s)/n_i$ and $d\Lambda$ by $d\hat\Lambda=dN_i(s)/Y_i(s)$, resulting in 
\begin{equation}\label{eq:covariance}
\widehat\cov(\hat\theta_{k,i},\hat\theta_{k',i})=\hat{a}_{k,i}\hat{a}_{k',i}\sum_{s \in \DD_i, s\leq t_{k}\wedge t_{k'}} \hat{H}_{k,i}(s)\hat{H}_{k',i}(s)\frac{1}{Y_i^2(s)} dN_i(s)
\end{equation}

We assume continuous survival distributions such that two events occur at the same time with probability zero. However, in actual applications event times are not measured precisely but they are typically rounded, e.g. to full days, such that tied event times may occur. To account for ties, the term $\frac{1}{Y_i^2(s)}dN_i(s)$ in (\ref{eq:covariance}) can be replaced by the sum
\begin{equation}\label{eq.tie_adjust}
    \sum_{j=0}^{dN_i(s)-1}\frac{1}{(Y_i(s) - j)^2} 
\end{equation} 
if $dN_i(s)\geq 1$, see e.g. Klein and Moeschberger\cite{klein2003survival}.

Let $\hat{\Sigma_i}$ denote the covariance matrix estimate for $(\hat\theta_{1,i},\hdots,\hat\theta_{m,i})$ with entries according to (\ref{eq:covariance}). Then the joint distribution of $(\hat\theta_1,\hdots,\hat\theta_m)$ can be approximated by a multivariate normal distribution with mean $(\theta_1,\hdots,\theta_m)$ and covariance matrix $\hat\Sigma=\hat\Sigma_0+\hat\Sigma_1$.

\subsection{Application to specific parameters}
\label{sec.estimates}
To quantify between-group differences, we consider a range of parameters:
differences in survival probabilities, differences in log survival probabilities, differences in cloglog transformed survival probabilities, differences in quantiles of the survival times, differences in log transformed quantiles, an average hazard ratio and the difference in restricted mean survival times. We also consider the Cox model score test (logrank test) statistic and (under the proportional hazard assumption) the hazard ratio corresponding to the Cox model.
For all these parameters,  estimators can be constructed that satisfy the assumptions laid out in Section \ref{sec.asym}. Especially, the estimators can be written in the form (\ref{eq:martingale_repr}), as will be detailed below. Hence, their joint distribution can be approximated by a multivariate normal distribution, with covariances estimated by equation (\ref{eq:covariance}). Based on this approximation, multiple hypotheses tests can be constructed.
The specific terms to calculate variance and covariance estimates according to (\ref{eq:covariance}) are summarized in Table \ref{tab.covar}.

\subsubsection{Survival probabilities}
The estimated difference in survival functions at times $t$ is 
$\hat\theta_j = \hat{S}_1(t)-\hat{S}_0(t)$.
The asymptotic properties and the martingale representation for survival function estimates are well established \cite{aalen2010history,fleming1991counting}. In particular,

\begin{equation}\label{eq:cumhazintegral}
\hat{\Lambda}_i(t)-\Lambda_i(t)= \int_0^t \frac{1}{Y_i(s)} dM_i(s)
\end{equation}
and by first order approximation \cite{fleming1991counting},
\begin{equation}\label{eq:Survintegral}
\hat{S}_i(t)-S_i(t) = -S_i(t)(\hat{\Lambda}_i(t)-\Lambda_i(t))+o_p(1/\sqrt{n_i})
\end{equation}
Hence the representation in terms of equation (\ref{eq:martingale_repr}) is
$\hat{S}_i(t)-S_i(t) \approx -S_i(t)\int_0^t \frac{1}{Y_i(s)} dM_i(s)$.

Alternatively, the ratio in survival probabilities at time $t$ may be of interest, which may be included in the proposed framework at the log-scale in terms of
$\hat\theta_j = \log\hat{S}_1(t)-\log\hat{S}_0(t)$.
Since $\log S(t) = -\Lambda(t)$ and by equation (\ref{eq:cumhazintegral}), the representation in terms of equation (\ref{eq:martingale_repr}) is
$\log\hat{S}_i(t)-\log S_i(t) \approx -\int_0^t \frac{1}{Y_i(s)} dM_i(s)$.
Estimates and confidence intervals for the log-ratio of survival probabilities may be backtransformed to obtain the respective quantities for the ratio $S_1(t)/S_0(t)$ at the original scale.

A further common transformation is the complementary log log (cloglog) of the estimated survival probability, such that the parameter
$\hat\theta_j = \log(-\log\hat{S}_1(t))-\log(-\log\hat{S}_0(t))$
may be included with the martingale representation
$\log(-\log\hat{S}_i(t))-\log(-\log S_i(t)) \approx -\frac{1}{\log S(t)}\int_0^t \frac{1}{Y_i(s)} dM_i(s)$.
Here, the transformed parameter $\exp(\hat\theta_j)$ may be of interest. This is an estimate for the ratio of cumulative hazards, $\Lambda_1(t)/\Lambda_0(t)$ (which under proportional hazards corresponds to the hazard ratio).

\subsubsection{Quantiles of the survival function}

The estimated between-group difference between the $\gamma$-quantiles of two survival time distributions is
$\hat\theta_j=\hat{q}_1(\gamma)-\hat{q}_0(\gamma)$.
A first order approximation is given by $\hat{q}_i(\gamma)-q_i(\gamma) = -\frac{\hat\Lambda_i(q_i(\gamma))-\Lambda_i(q_i(\gamma))}{\lambda_i(q_i(\gamma))}+o_p(1/\sqrt{n_i})$ \cite{sander1975weak,sander1975asymptotic}. Again using (\ref{eq:cumhazintegral}) results in the representation
\[\hat{q}_i(\gamma)-q_i(\gamma) \approx \frac{-1}{\hat\lambda_i(q_i(\gamma))}\int_0^t \frac{1}{Y_i(s)}dM(s)\].

Here a consistent estimate $\hat\lambda_i(q_i(\gamma))$ for the hazard at the respective quantile is required. We use a relatively simple approach and estimate $\hat\lambda_i(q_i(\gamma))$ under a locally constant hazard approximation as follows: Let $e_i$ be the total number of observed events in group $i$. Define a positive finite constant bandwidth $K$ and the boundaries for a time interval that contains at least $K\sqrt{e_i}$ events as

$t_{low}=\max\{0,\max\{t \in \DD_i: N_i(t) \leq N_i(\hat q_{\gamma,i})-K\sqrt{e_i}\}\}$
and
$t_{up}=\min\{\max\{t \in \DD_i\},\min\{t \in \DD_i: N_i(t) \geq N_i(\hat q_{\gamma,i})+K\sqrt{e_i})\}\}$.
Here, the maximum and minimum of an empty set are defined as $-\infty$ and $\infty$, respectively. 
The hazard in the interval is estimated as ratio of the number of events and the sum of all observed times, i.e. $\hat\lambda_i(q_i(\gamma))=\frac{N_i(t_{up})-N_i(t_{low})}{\int_{t_{low}}^{t_{up}} Y_i(s) ds}$.

With increasing number of events, the interval becomes narrower at while the absolute number of events within the interval gets larger. Under the assumption of continuous hazard function and by consistency of $\hat q_i(\gamma)$, the resulting estimate is consistent. In the actual calculations we used $K=2$. 
Alternatively, the local hazard may be estimated by
kernel-density estimation \cite{muller1994hazard}. 

When the ratio of quantiles is of interest, the difference at the log scale may be defined as parameter of interest such that
$\hat\theta_j=\log\hat{q}_1(\gamma)-\log\hat{q}_0(\gamma)$.
By application of the delta method to the original approximation we obtain the required presentation as
\[\log\hat{q}_i(\gamma)-\log q_i(\gamma) \approx \frac{-1}{\hat{q}_i(\gamma) \hat\lambda_i(q_i(\gamma))}\int_0^t \frac{1}{Y_i(s)}dM(s)\]

\subsubsection{Average hazard ratio}
A general class of average hazard ratios can be defined as $\frac{\int_0^L W(s) d\hat{\Lambda}_1(s) }{\int_0^L W(s) d\hat{\Lambda}_0(s) }$, where $L$ is a predefined time point and $W(s),s\geq 0$ is a non-negative monotonically decreasing weight function with values in $[0,1]$  \cite{kalbfleisch1981estimation}. We consider the average hazard ratio with weight function $W(s)= \hat{S}_0(s)\hat{S}_1(s)$ and its corresponding estimate $\hat{W}(s)=\hat{S}^{-}_0(s)\hat{S}^{-}_1(s)$. Here the left continuous estimator of the survival function is used to obtain a predictable function, which is required in the eventual application of the martingale central limit theorem. Note that both $\hat{S}$ and $\hat{S}^{-}$ are uniformly consistent estimates of $S$, and with sufficient sample size their numeric difference is negligible. 

The average hazard ratio based on the considered weight function is identical to $\frac{P(T_1 \wedge L >T_0 \wedge L)}{P(T_1 \wedge L< T_0 \wedge L)}$ and can be interpreted as a non-parametric effect measure similar to a Mann-Whitney statistic. Unlike the Cox model hazard ratio estimate, the limiting value of this average hazard ratio estimate does not depend on the censoring distribution (also see simulation scenarios 2 and 3 in Section \ref{sec.sim}).

To embed the average hazard ratio in the proposed framework we utilize  the log transformed estimate
$\hat\theta_j=\log\left(\int_0^L \hat{W}(s) d\hat{\Lambda}_1(s) \right)-\log\left(\int_0^L \hat{W}(s) d\hat{\Lambda}_0(s) \right)$.

The contribution of group $i$ to the estimate of the log-average hazard ratio immediately fits into the proposed framework as
$\int_0^L \hat{W}(s) d\hat{\Lambda}_i(s) - \int_0^L \hat{W}(s) d\Lambda_i(s)=\int_0^L \hat{W}(s) \frac{1}{Y_i(s)}dM_i(s)$.

By application of the delta method the representation according to (\ref{eq:martingale_repr}) for the log transformed term is
\begin{eqnarray*}
\log\left(\int_0^L \hat{W}(s) d\hat{\Lambda}_i(s)\right)  - \log\left(\int_0^L \hat{W}(s) d\Lambda_i(s) \right)\approx \\ \frac{1}{\int_0^L \hat{W}(s) d\hat{\Lambda}_i(s) } \int_0^L \hat{W}(s) \frac{1}{Y_i(s)}dM_i(s) 
\end{eqnarray*}
Note that the statistic $\sqrt{n_i}\int_0^L \hat{W}(s) \frac{1}{Y_i(s)}dM_i(s)$ is asymptotically equivalent to $\sqrt{n_i}\int_0^L W(s) \frac{1}{Y_i(s)}dM_i(s)$. This follows from the following arguments:
By the martingale central limit theorem, the difference between both expressions, $\sqrt{n_i}\int_0^L \hat{W}(s) \frac{1}{Y_i(s)}dM_i(s) - \sqrt{n_i}\int_0^L W(s) \frac{1}{Y_i(s)}dM_i(s)=\sqrt{n_i}\int_0^L (\hat{W}(s)-W(s)) \frac{1}{Y_i(s)}dM_i(s)$, converges in distribution to a normal distribution with mean zero and variance $\int_0^L (\hat{W}(s)-W(s))^2 \frac{n}{Y(s)} d\Lambda(s)$.
Because $\hat{S}_i^{-}$ and, consequently, $\hat{W}(s)$ are uniformly stronlgy consistent estimators, the variance of $\sqrt{n_i}\int_0^L (\hat{W}(s)-W(s)) \frac{1}{Y_i(s)}dM_i(s)$ can be bounded by $\int_0^L (\hat{W}(s)-W(s))^2 \frac{n}{Y(s)} d\Lambda(s) \leq \max_{0\leq t\leq L}((\hat{W}(s)-W(s))^2) \int_0^L \frac{n_i}{Y(s)} d\Lambda(s)$.
Since $\max_{0\leq t\leq L}((\hat{W}(s)-W(s))^2)$ converges to 0 a.s. 
and $\int_0^L \frac{n_i}{Y(s)} d\Lambda(s)$ 
converges to a constant (the asymptotic variance of 
$\hat{\Lambda}_i(L)$), the variance of 
$\sqrt{n_i}\int_0^L (\hat{W}(s)-W(s)) \frac{1}{Y_i(s)}dM_i(s)$ converges to 0.
Hence, the asymptotic arguments established in Section \ref{sec.asym} are not affected by using estimated weights $\hat{W}$ instead of the true (unknown) weights $W$ and $\int_0^L \hat{W}(s) \frac{1}{Y_0(s)}dM_0(s)$ and $\int_0^L \hat{W}(s) \frac{1}{Y_1(s)}dM_1(s)$ are asymptotically uncorrelated.

\subsubsection{Restricted mean survival time (RMST)}
The RMST in group $i$ up to a pre-specified time-point $L$ is $\mu_i=\int_0^L S_i(t)dt$. Let $D_i$ be the number of unique event times $t_{i,1} < \hdots < t_{i,D_i-1} \leq L$ in group $i$ that are less or equal $L$. Further define $t_{i,0}=0$ and $t_{i,D_i+1}=L$ and $\Delta t_{i,j}= t_{i,j+1}-t_{i,j}$. The according estimate for the RMST is $\hat\mu_i=\sum_{j=0}^{D_i} \hat S_i(t_j)\Delta t_{i,j}$ and the estimated RMST difference between the two groups is
$\hat\theta_j=\hat\mu_1-\hat\mu_0$.
$\hat\mu_i$ may be represented in terms of equation (\ref{eq:martingale_repr}) \cite{hasegawa2020restricted,zhao2012utilizing}. First note that 
\begin{eqnarray*}
\hat \mu_i - \mu_i = 
\sum_{j=0}^{D_i} \hat S_i(t_j)\Delta t_{i,j} - \sum_{j=0}^{D_i} S_i(t_j)\Delta t_{i,j} + \sum_{j=0}^{D_i} S_i(t_j)\Delta t_{i,j} - \int_0^L S_i(t)dt=\\
\sum_{j=0}^{D_i} \Delta t_{i,j}(\hat S_i(t_j)-S(t_j)) + o_p(1/\sqrt{n_i})
\end{eqnarray*}
where we use that the error of the integral approximation is of order $1/n$.
By replacing $\hat S_i(t_j)-S_i(t_j)$ by equation (\ref{eq:Survintegral}) and by changing the order of integration and summation the required form is obtained as
\[
\hat \mu_i - \mu_i \approx -\int_0^L \left(\sum_{j \in \{1,...,D_i:t_j \geq s\}} \Delta t_j S_i(t_j)\right) \frac{1}{Y_i(s)}dM_i(s) 
\]

\subsubsection{Cox model score test (logrank test)}
The logrank test for the null hypothesis $H_0:\lambda_0(s)=\lambda_1(s), \forall 0 \leq 0 \leq L$ may be of interest also under non-proportional hazard settings. The usual logrank test is asymptotically equivalent to the Cox model score test for the null hypothesis $\beta = 0$, where $\beta$ is the Cox model hazard ratio. The tests are equivalent up to the variance estimate for the score test statistic. The score test may be directly included in the proposed framework to adjust for multiple testing as shown below.

For subject $j= 1,\hdots,n_i$ in group $i \in \{0,1\}$, let $y_{ij}(t)\in \{0,1\}$ indicate whether the subject is at risk at time $t$, and let $N_{ij}(t) \in \{0,1\}$ be the number of events of the subject in the time interval $[0,t]$. Let $M_{ij}(t)=N_{ij}(t)-y_{ij}(t)\Lambda_i(t)$.

In a Cox model comparing two treatment groups up to a time point $L$, the score function (i.e. the derivative of the log partial likelihood) is
\begin{equation}\label{eq.coxscore}
U(\beta)=\sum_{i=0}^1\sum_{j=1}^{n_i} \int_0^L \left(i-\frac{Y_1(s)\exp(\beta)}{Y_0(s)+Y_1(s)\exp(\beta)}\right)dN_{ij}(s)
\end{equation}
It can be shown  \cite{gill1984understanding} that under the proportional hazards assumption (i.e. $\lambda_1(t)=\lambda_0(t)\exp(\beta), \forall t\geq0$), and for $\beta=\beta_0$ being the true parameter value, $dN_{ij}$ may be replaced by $dM_{ij}$, such that
\begin{equation}\label{eq.coxscore_dM}
U(\beta_0)=\sum_{i=0}^1\sum_{j=1}^{n_i} \int_0^L \left(i-\frac{Y_1(s)\exp(\beta_0)}{Y_0(s)+Y_1(s)\exp(\beta_0)}\right)dM_{ij}(s)
\end{equation}
which can be rewritten as
\begin{equation}\label{eq.scoretestU}
        U(\beta_0)= \int_0^L
        \frac{Y_0(s)Y_1(s)}{Y_0(s)+Y_1(s)\exp(\beta_0)} \frac{1}{Y_1(s)}dM_1(s) - 
        \int_0^{L} 
        \frac{Y_0(s)Y_1(s)\exp(\beta_0)}{Y_0(s)+Y_1(s)\exp(\beta_0)} \frac{1}{Y_0(s)}dM_0(s)
\end{equation}
Note that under the null hypothesis $H_0:\lambda_0(s)=\lambda_1(s), \forall s \leq 0 \leq L$, the proportional hazard assumption holds. Further note that rejection of $\beta=0$ entails rejection of $\lambda_0(s)=\lambda_1(s)$ at least for some $s$. The reverse is not necessarily true, which results in low power of the logrank test under crossing hazards.
The test statistic for the score test of the null-hyptothesis $\beta=\beta_0$ is $U(\beta_0)$ as defined in \ref{eq.scoretestU}.
Hence, the Cox model score test for the null hypothesis $\beta=0$ can be included in the framework of Section \ref{sec.asym} by defining a parameter estimate $\hat\theta_j = U(0)$ and by setting $H_{k,i}(s) = \frac{Y_0(s)Y_1(s)}{Y_0(s)+Y_1(s)}$ and $a_{k,i}=1$ in equation (\ref{eq:martingale_repr}). 

Note that Assumption 3 about uncorrelated contributions from both groups is satisfied for the score test despite both terms in equation (\ref{eq.scoretestU}) contain the at risk process of both groups, because we assume that the probability for equal event times is 0 (and ties only occur due to rounding of observed event times). Under this assumption, theorems 2.5 §2 and 2.4 §4 of Fleming and Harrington show that the covariance of statistics of the type assumed in equation (\ref{eq:martingale_repr}) is 0.

\subsubsection{Cox model hazard ratio}
The Cox model hazard ratio can be included in the proposed framework of simultaneous inference under the assumption of proportional hazards.
First note that the estimate of the log hazard ratio, $\hat\beta$, is the solution of $U(\hat\beta)=0$.
Next, by standard asymptotic results and with $\beta_0$ the true log hazard ratio
$
\hat\beta-\beta_0 \approx -\left(\frac{dU}{d\beta}\right)^{-1}(\beta_0)U(\beta_0)
$ \cite{andersen1982cox}.
Here, $U(\beta_0)$ can be decomposed in contributions from the two groups according to (\ref{eq.scoretestU}).
The Hessian matrix $\mathcal{H}=\frac{dU}{d\beta}$ can be estimated by
\[
    \hat{\mathcal{H}}=-\sum_{i=0}^1 \int_0^L \frac{Y_0(s)Y_1(s)\exp(\hat\beta)}{Y_0(s)+Y_1(s)\exp(\hat\beta)}dN_i(s)
\]
Therefore, the Cox model hazard ratio can be included in the framework of Section \ref{sec.asym}, by setting $\theta_j=\hat\beta$,
$H_{k,i}(s) = \frac{Y_0(s)Y_1(s)\exp(i\hat\beta)}{Y_0(s)+Y_1(s)\exp(\hat\beta)}$ and $a_{k,i}= (-\hat{\mathcal{H}})^{-1}$. The theory of Section \ref{sec.asym} applies to the hazard ratio under the assumption of proportional hazards. The operating characteristics of simultaneous inference including the hazard ratio under non-proportional hazards are explored in the Supplemental material.

\subsection{Resampling based covariance matrix estimate}
\label{sec.perturb}
As an alternative method, the covariance matrix for a vector of parameter estimates $\hat\btheta=(\hat\theta_1,\hdots,\hat\theta_m)$ may be estimated by a resampling approach similar to perturbation methods
described by Zhao et al.\cite{zhao2012utilizing,zhao2016restricted} and Parzen et al.\cite{parzen1997simultaneous}, which can be regarded as parametric bootstrap.

Based on the martingale central limit theorem and equation (\ref{eq:covariance}), the process $\hat \Lambda_i(t)-\Lambda_i(t)=\int_0^t \frac{1}{Y_i(s)}dM_i(s), t \geq 0$ can be approximated by a Gaussian process with independent increments and covariance function $\cov (\hat \Lambda_i(t)-\Lambda_i(t),\hat \Lambda_i(t')-\Lambda_i(t'))=var(\hat \Lambda_i(t)-\Lambda_i(t))=\int_0^t \frac{dN_i(s)}{Y_i^2(s)}$. Accordingly, the distribution of this process can be approximated by the distribution of the process $P(t)= \sum_{j=1}^{n_i} \int_0^t Z_j \sqrt{\frac{dN_{ij}(s)}{Y_i^2(s}},t \geq 0$, where $Z_j,j=1,\hdots,n_i$ are independent standard normal random variables. 

A perturbation sample of $\hat\Lambda-\Lambda$ is defined as $P^*(t)= \sum_{j=1}^{n_i} \int_0^t z_j \sqrt{\frac{dN_{ij}(s)}{Y_i^2(s}},t \geq 0$, where $z_j,j=1,\hdots,n_i$ are realisations of independent standard normal random variables. And a perturbation of $\hat\Lambda$ is defined as $\hat\Lambda^*=\hat\Lambda+P^*$. Similar as with equation \ref{eq:covariance}, in case of ties the expression $\frac{dN_{ij}(s)}{Y_i^2(s}$ may be replaced by (\ref{eq.tie_adjust}) if $dN_i(s)\geq 1$.

A vector of parameter estimates can be regarded as function  $\hat\btheta(\hat\Lambda_0,\hat\Lambda_1)$ of the estimated cumulative hazard functions.
In the proposed perturbation approach, the distribution of $\hat\btheta(\hat\Lambda_0,\hat\Lambda_1)$ given the true cumulative hazard functions $\Lambda_0$ and $\Lambda_1$ is approximated by the distribution of $\hat\btheta(\hat\Lambda_0^*,\hat\Lambda_1^*)$ given $\hat\Lambda_0$ and $\hat\Lambda_1$.

To estimate the covariance matrix of $\hat\btheta(\hat\Lambda_0,\hat\Lambda_1)$, a large number of $K$ perturbation pairs $(\hat\Lambda_0^*,\hat\Lambda_1^*)_l,l=1,\hdots,K,$ is generated and for each pair the estimate $\hat\btheta^*_l=\hat\btheta((\hat\Lambda_0^*,\hat\Lambda_1^*)_l),$ is calculated. Now consider the matrix $(\hat\btheta^*_1,\hdots,\hat\btheta^*_K)$ and let $\hat\bTheta^*$ be the corresponding row-mean centered matrix. Then the covariance matrix of $\hat\btheta$ can be estimated by the empirical covariance matrix as 
\begin{equation}\label{eq.covar_perturb}
\hat\cov_{pert}(\hat\btheta)=\hat\bTheta^* \hat\bTheta^{*T} / (K-1)
\end{equation}

\section{Simultaneous inference}\label{sec:inference}

\subsection{Maximum-type hypothesis tests for multivariate normal statistics}
\label{sec.test}
We aim to test hypothesis with respect to parameters $\btheta=(\theta_1,\hdots,\theta_m)$, as defined in the previous sections, with strong control of the family-wise type I error rate. We consider one-sided elementary null hypotheses $H_k:\theta_k \leq \theta_k^{(0)}$, $k=1,\ldots,m$ and the global intersection null hypothesis of no difference in any of the parameters $H_0 = \cap_{k=1}^m H_k$.  In superiority trials, depending whether the parameter of interest are ratios or differences between treatment and control, the $\theta_k^{(0)}$ is set to 1 or 0, respectively. The considered hypothesis tests can also be defined for two-sided hypotheses, but for the purpose to show superiority of treatment over a control we regard one-sided tests to be more relevant.

Multiple hypothesis tests can be constructed based on the multivariate normal approximation of the estimates $\hat\btheta=(\hat\theta_1,\hdots,\hat\theta_m)$, see e.g. \cite{hothorn2008simultaneous}. Correlations of estimates can be high, hence the multiplicity correction based on the approximate joint distribution can be moderate compared to methods that control the  family-wise type I error rate whithout taking into account the correlation structure, as the Bonferroni correction.
We consider maximum type tests. Other test statistics, such as sum or chi-squared type statistics may also be used within the multivariate normal framework. Consider a vector of $m$ parameter estimates $\hat\btheta \asim N_m(\btheta,\hat\bSigma)$.
Let $\hat\bV$ be the diagonal matrix with diagonal entries of $\hat\bSigma$
Define a vector of elementary test statistics $\bT=\hat \bV^{-1/2}(\hat\btheta-\btheta_0)$. Under the global null hypothesis $H_0:\btheta=\btheta_0, \bT \asim N(0,\hat\bV^{-1/2}\hat\bSigma \hat\bV^{-1/2})$ 
A one-sided p-value for the test of $H_0$ is given by $P\{\max(\bZ)\geq \max(\bT)\}$, where $\bZ\sim N_m(0,\hat \bV^{-1/2}\hat\bSigma \hat \bV^{-1/2})$.\\

A single step maximum-type test for elementary null hypothesis can then be defined via multiplicity adjusted p-values for the  hypotheses $H_k,k \in \{1,\hdots,m\}$ and is given by $P\{\max(\bZ)\geq T_j\}$.

The single step test can be improved applying the closed testing procedure \cite{marcus1976closed} where each intersection null hypothesis $\cap_{k \in I} H_k, I \subset {1,\hdots,m}$ is tested by a maximum-type test for intersection hypotheses. The elementary hypothesis $H_k$ is then rejected by the closed test, if all intersection hypotheses that include $H_k$ (i.e. all $\cap_{k \in I} H_k$ with $I \subset {1,\hdots,m}$ and $k \in I$) are rejected with the corresponding maximum-type test. A multiplicity adjusted p-value is given by the maximum of the elementary p-values of these intersection hypotheses.

\subsection{Simultaneous confidence intervals}\label{sec.ci}
Lower confidence bounds with simultaneous coverage probability $1-\alpha$ that correspond to the inversion of a one-sided single-step multiple testing procedure are given by 
$\hat\theta_j - \hat V_j^{1/2}q_\alpha$,
where $q_\alpha$ is defined as $q_\alpha:P(\max(\bZ) \geq q_\alpha)=\alpha$. The boundaries of two-sided confidence intervals with simultaneous coverage probability $1-\alpha$ that correspond to the inversion of a two-sided single-step test are defined as $\hat\theta_j \pm \hat V_j^{1/2}q'_\alpha$, where $q'_\alpha$ is defined as $q'_\alpha:P(\max(|\bZ|) \geq q')=\alpha$.

In an actual analysis, the closed testing procedure may be preferred over the single step procedure as it is more powerful. It is possible to construct confidence regions that correspond to the inversion of the closed testing procedure. However, such regions would typically be of complex shape and not easily interpreted, and the projections of such a region onto univariate confidence intervals would typically not retain the advantage of the closed test over the single step test \cite{strassburger2008compatible}.

\section{Simulation study} \label{sec.sim}

In six different simulation scenarios and for different parameter sets, we studied the 
operating characteristics of the proposed multivariate inference methods and compared them with the Bonferroni-Holm multiplicity adjustment and a single logrank test. 

We considered seven parameter sets. In parameter set 1, we included the average hazard ratio and the restricted mean survival time as an example of joint inference using two alternative summary measures of a possible treatment benefit.

In parameter set 2, we included the between-group difference in survival probabilities after one, two and three years as well as  25\% quantiles and the restricted mean survival time. This parameter set was intended to illustrate an 
analysis where one summary measure (RMST) is combined with several statistics that allow a point-wise description of the survival differnces.
Parameter set 3 was defined similarly, albeit replacing survival probabilites and quantiles by their log-transformations. This set was included to study a possible effect of log-transformation on the finite sample properties of the distributional approximation. In similar spirit, parameter set 4 included cloglog transformed survival probabilites combined with the average hazard ratio.

Parameter sets 5 to 7 were designed to study simultaneous inference for the logrank test combined with several quantifying parameters. Parameter sets 5 and 6 included the logrank test and differences in untransformed or cloglog transformed survival probabilites, respectively. In parameter set 7 we combined the logrank test with estimates for the average hazard ratio and restricted mean survivial time.

In all scenarios, a cut-off value of three years was set for the restricted mean survival time, for (average) hazard ratios and for the calculation of the logrank test.

Table \ref{tab:parametersets} shows a summary of the considered parameter sets and the true values of the respective parameters in the simulation scenarios. Further simulation scenarios, including also the Cox model hazard ratio (despite its limitation) are considered in the Supplemental material.

To assess the coverage probabilities of confidence intervals, simulations were performed for parameter sets 1 to 4 (see Table \ref{tab:parametersets}) with sample sizes 200, 500 or 1000 per group. Adjusted and unadjusted two-sided confidence intervals with nominal (simultaneous) coverage probability of 95\% were calculated. The asymptotic covariance matrix estimate and the perturbation based estimate, both with adjustment for ties, were used.

To assess type I error rate and power of hypothesis tests, simulations were performed for parameter sets 5 to 7 (see Table \ref{tab:parametersets}) with sample size 200 per group, under the alternative and under the null hypothesis of identical survival curves. In the latter case, data for both groups was sampled from the distribution of the control group. 
For each parameter set $(\theta_1,\hdots,\theta_m)$, the elementary null hypotheses $H_1:\theta_1=0,\hdots,H_m:\theta_m=0$ versus one-sided alternatives were considered, as well as the global null hypothesis $\cap_{i=1}^m H_i$. The null hypotheses were tested with the closed testing procedure described in Section \ref{sec:inference} and, for comparison, with unadjusted and with Bonferromi-Holm adjusted tests where the elementary p-values were computed based on univariate normal approximations. The nominal (family-wise) one-sided significance level was set to 0.025. To calculate the critical values the asymptotic covariance matrix estimate with adjustment for ties was used (see Section \ref{sec.asym}). Simulation results using the resampling based covariance estimate (see Section \ref{sec.perturb}) are given in the Supplemental Material.

\subsection{Simulation scenarios} \label{sec.scen}

In all scenarios we compare two groups with equal number of patients per group in $(200,500,1000)$. Recruitment was assumed to be uniform over 1 or 1.5 years (depending on the scenario) and the maximum follow up time was 3.5 years in all scenarios.
Furthermore, we applied random censoring according to an exponential distribution with rate $-\log(0.9)$ such that, given no other events occur, on average within one year 10\%  and within 2 years 19\% of subjects are censored. Simulated survival times were rounded to full days to reflect the degree of precision and the occurrence of ties as observed in actual trials. All simulations were repeated 50,000 times.

\begin{figure}[ht]
    \begin{center}
          \includegraphics[scale=.49,clip,trim=0cm 0.4cm 0.00cm 0.5cm]{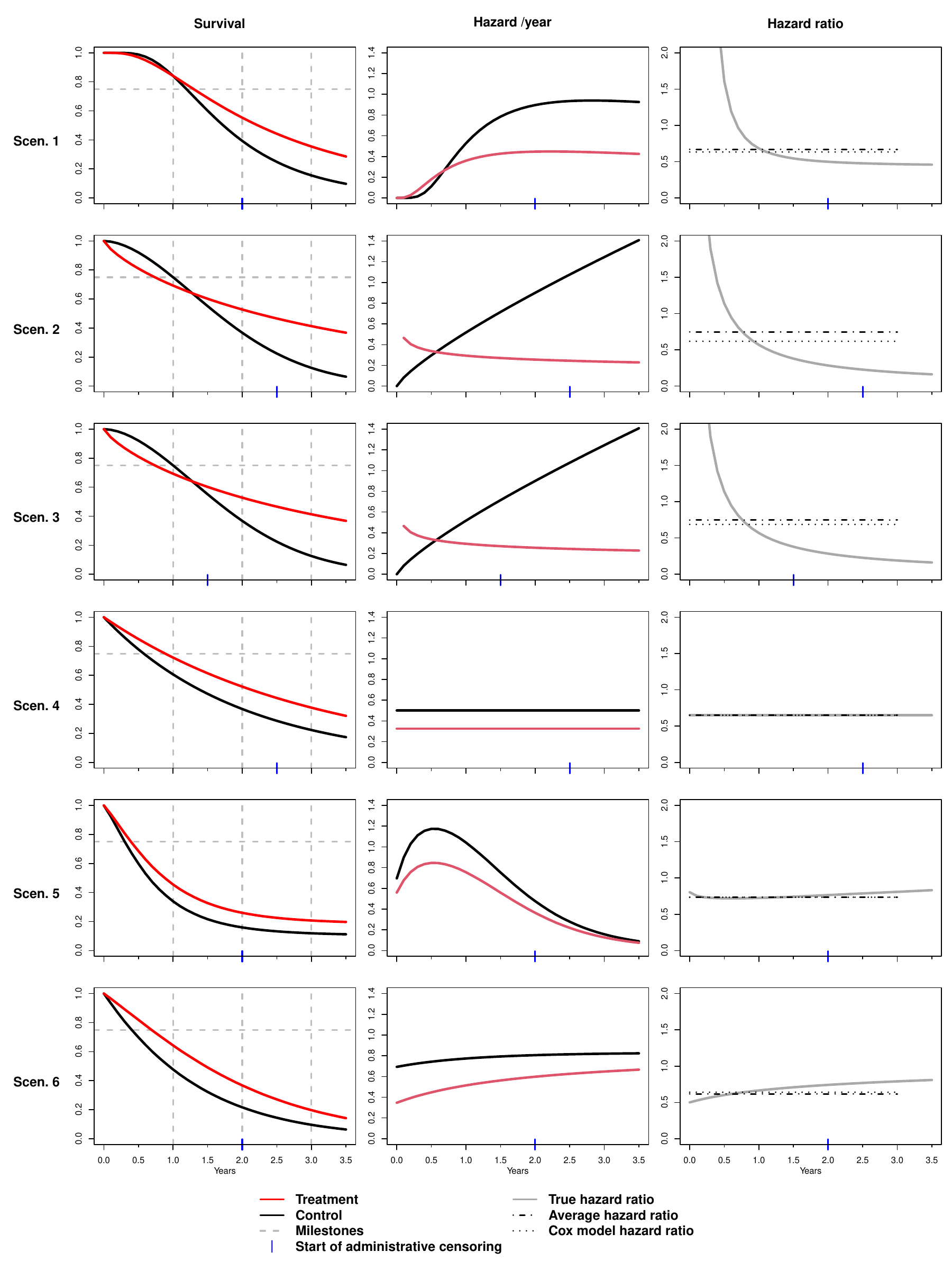}
    \end{center}
\caption{Simulation scenarios. Left column: True survival functions for treatment and control group. Vertical dashed lines indicate the 1, 2 and 3-year time-point. The horizontal line indicates the 25\% quantile. Middle column: True hazard rates (per year) as function of time for treatment and control group. Right column: Hazard ratio between treatment and control as function of time. Dotted and dash-dotted lines indicate the large sample limit of the Cox model hazard ratio estimate and the true average hazard ratio as defined in Section  \ref{sec.estimates}, both calculated with a cut-off at three years.}
\label{fig.scenarios}
\end{figure}

The scenarios are described in detail below, and the resulting survival functions, hazard functions and hazard ratios as function of time are shown in Figure \ref{fig.scenarios}.

\textbf{Scenario 1, Delayed onset of treatment effect:} In Scenario 1 we sampled data for the treatment group from a lognormal$(0.8,0.8^2)$ distribution and data for the control group from a lognormal$(\log(\exp(0.8)-0.5),(\log(\exp(0.8)-0.5))^2)$ distribution. 
The distributions were chosen to resemble a setting with delayed onset of the treatment effect. During the inital phase, the control group has slightly better survival, which may occur if the treatment effect is observed only after a certain duration of treatment, but the potential risk of side effects increases immediately after treatment start. 
The recruitment phase in the simulation was 1.5 years.

\textbf{Scenario 2, Crossing survival curves, fast recruitment:}
In Scenario 2 we used Weibull(2,1.8) and Weibull(3.5,0.8) distributions (where parameters refer to scale and shape) for the treatment and the control group, respectively. The resulting distributions, as illustrated in Figure \ref{fig.scenarios}, show a pronounced crossing of the survival functions and the hazard functions. The assumed duration of the recruitment phase was one year.

\textbf{Scenario 3, Crossing survival curves, slow recruitment:}
Scenario 3 was identical to scenario 2, except that the duration of the recruitment phase was two years, which results in a modified censoring pattern.

\textbf{Scenario 4, Proportional hazards:}
In Scenario 4, data are sampled from Exp(0.5) and Exp($0.5*0.65$) distributions, meeting the proportional hazards assumption with a hazard ratio of 0.65, with a recruitment phase of one year.

\textbf{Scenario 5, Cure fraction:}
In scenario 5 we assumed that 30\% of patients belong to a subpopulation of strong treatment responders in whom the treatment leads to complete cure and we further assumed that in the control group, patients are switched to active treatment after disease progression with 70\% probability. Transitions to progression or death were assumed to be governed by independent processes with constant rates. Hazard rates for death were 0.69 per year before progression and 2.77 per year after progression and the progression rate was 1.39 per year. (This corresponds to median event times for the three processes of 12, 6 and 3 months, respectively.) The recruitment phase was 1.5 years. Data simulation for scenarios 5 and 6 was performed using The R library nph \cite{ristl2021delayed}.

\textbf{Scenario 6, Rescue medication:}
In the final scenario we assume that a rescue medication is applied to all patients in both groups after disease progression. The assumed hazard rates per year for death were 0.35 under treatment, 0.69 under control and 0.83 under rescue medication (corresponding to median times of 24, 12 and 10 months). Progression rates were 0.52 under treatment and 0.92 under control (corresponding to median times of 16 and 9 months). The recruitment phase was 1.5 years.

\subsection{Simulation results}
\begin{figure}[!p]
    \begin{center}
          \includegraphics[page=1,scale=.6,clip,trim=0cm 29cm 0.00cm 0cm]{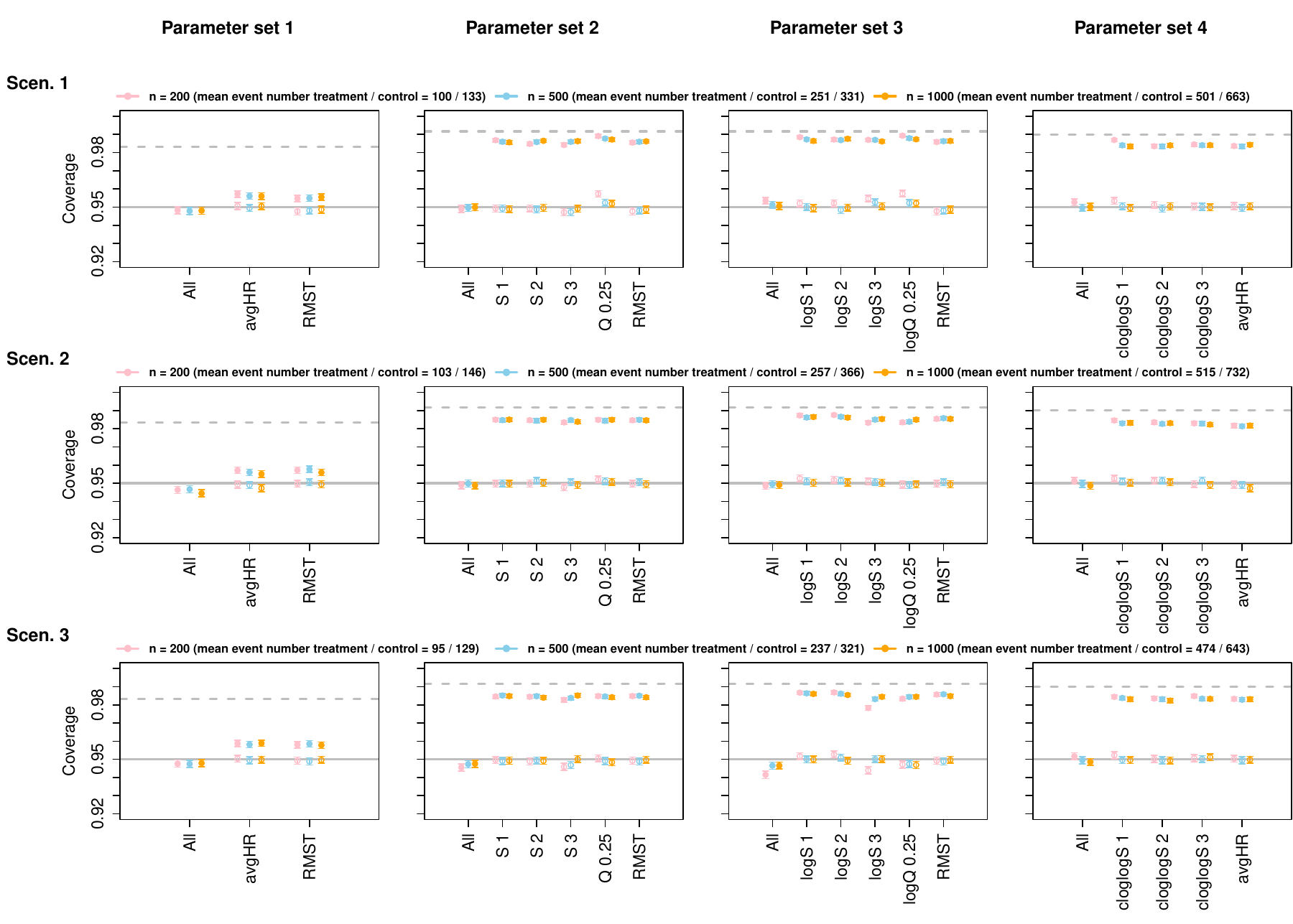} 
    \end{center}
\caption{Empirical coverage of confidence intervals for Scenarios 1-3, based on the multivariate normal distribution with asymptotic covariance matrix estimate. Filled circles show the simultaneous coverage (All) and univariate coverage probability of multiplicity adjusted intervals for single parameters (abbreviations as in Table \ref{tab:parametersets}). Open circles represent the coverage of unadjusted univariate confidence intervals. Error bars represent 95\% Wald confidence intervals for the respective coverage probabilities. For comparison, the horizontal solid line indicates the nominal coverage of 95\%, and the horizontal dashed line indicates the univariate confidence level that would result from a Bonferroni adjustment for the respective number of parameters.
}
\label{fig.coverage_asym123}
\end{figure}

The empirical coverage probabilities of simultaneous confidence intervals as well as of unadjusted, univariate confidence intervals were in general close to the nominal 95\%, for both the asymptotic and the perturbation covariance matrix estimate. In some scenarios, coverage probabilities showed a deviation in the order of one percentage point from the nominal level when the smallest studied sample size of 200 subjects per group was used. Results for the asymptotic covariance estimate are shown in Figures \ref{fig.coverage_asym123} and  \ref{fig.coverage_asym456}, results for the perturbation based covariance estimate are shown in the supplemental material in Figures S1 and S2.
The univariate coverage of multiplicity adjusted intervals was greater than 95\%, but the intervals were less conservative than simple Bonferroni adjusted intervals. Similar results in general close to nominal coverage were observed for parameter sets including the Cox model hazard ratio, see supplemental Figures S3 and S4.

Intervals based on the asymptotic and on the perturbation covariance matrix estimate performed largely similar, with some exceptions: In scenario 2 (crossing hazards) the simultaneous coverage for the parameter set containing average hazard ratio and RMST difference was slightly liberal even with large sample size when using the asymptotic covariance estimate. With the perturbation approach the coverage was conservative with small sample size and almost perfectly matching the nominal level with large sample size. Intervals for the difference in cloglog transformed survival were overly conservative with the perturbation approach applied to small sample sizes, in particular for early time-points. These intervals had coverage close to the nominal level with the asymptotic covariance estimate.

The type I error rate of one-sided unadjusted hypothesis tests for the studied parameters was well controlled at the 2.5\% level with the excpetion of tests for untransformed survival probabilites. However, it is well known that the normal approximation for untransformed survival probabilites may not be entirely appropriate with small sample sizes. In contrast, tests for cloglog transformed survival did control the type I error rate. See column 'T1E unadj' in Tables \ref{tab.hyp_test_1} and \ref{tab.hyp_test_2}.

Without adjustment, the type I error rate for the global null hypothesis of no difference in any included parameter was in the order of 6.5\% to 7.5\% in the studied scenarios. Multiplicity adjustment using the proposed multivariate normal approximation resulted in type I error rates close to the nominal 2.5\%. Some inflation was still observed for parameter sets containing untransformed survival probabilities, which likely results from the inaccuracy even of the univariate approximation for these parameters (see column 'T1E adj' in Tables \ref{tab.hyp_test_1} and \ref{tab.hyp_test_2}). 
Adjustment using the Bonferroni-Holm test resulted in strictly conservative tests and adjusted significance levels below those of the multivariate normal-based closed test (see column 'T1E adj' and 'T1E Holm' in the result tables).

The power of the multivariate normal-adjusted tests for the global null hypothesis was on average 4.0 percentage points below the power of corresponding unadjusted tests. For comparison, Bonferroni adjusted tests were on average 7.7 percentage points less powerful than the unadjusted tests. (See columns regarding power and rows with parameter label 'Any' Tables \ref{tab.hyp_test_1} and \ref{tab.hyp_test_2}.) Similarly, the power for elementary hypothesis tests was on average 4.4 percentage points lower with the multivariate normal-based closed test compared to unadjusted univariate tests. The Bonferroni-Holm procedure resulted on average in 7.2 percentage points lower power compared to unadjusted tests.

That means, averaged across the studied settings, the proposed testing procedure retains almost half the power loss, which would occur with a simpler Bonferroni-Holm approach. 

When comparing the approach to test multiple parameters with a single Cox model score test, in scenarios with strong non-proportionality of hazard functions (scenarios 1, 2 and 3), the hypothesis test for a difference in 3-year survival or 3-year cloglog-transformed survival was of similar (scenario 1) or larger power (scenarios 2 and 3) compared to the score test. 

When including the difference for 1-year, 2-year, 3-year survival (or cloglog transformed survival) and the score test in one parameter set and adjusting for multiple testing, the power to show a difference in at least one considered parameter was similar (scenario 1) or considerably larger (Scenarios 2 and 3) than the power of a single unadjusted score test. Further, the power to show a difference in at least one parameter under multiplicity adjustment was similar to the unadjusted power of the most powerful univariate comparison (either 3-year survival or score test).

This implies that, first, under severely non-proportional hazards, testing for differences at a well chosen milestone time-point can be more powerful than the score test. And, secondly, testing several milestone time-points and adjust for multiplicity will often be a better choice than selecting one time-point in advance and avoid multiple testing, as the multiplicity adjustment with the proposed method will mostly maintain the power of the most powerful univariate comparison.

When comparing the score test for the Cox model hazard ratio with tests for the two other summary measures (average hazard ratio and RMST difference), the score test performs by far best in the three scenarios with strongly non-proportional hazards. 

In scenarios with proportional hazards (scenario 4) or moderate non-proportionality (scenarios 5 and 6), the score test was more powerful, in the order of 10 percantage points, than the best comparison for survival at either milestone time-point. Also, the unadjusted power of the score test was larger than the adjusted power to reject at least one of the considered null hypotheses by two to five percantage points. Tests for average hazard ratio and RMST difference had almost identical power as the score test in scenarios 4 and 6 and when combining all three tests, the adjusted power to reject at least one null hypothesis was also at the almost same value as the power of the unadjusted tests. In scenario 6, the score test and the test for RMST difference both had a power of approximately 77\%, whereas the test for average hazard ratio was at 71.6\%. Still, the power of the better tests was almost retained in adjusted power to reject at least one null hypothesis with a value of 75\%.

Taken together, the simulation results show that in terms of power to show at least some difference, the score test (or logrank test) is in most settings a robust choice. However, to increase robustness or to aid in the interpretation of the pattern of differences between survival curves, the test may be complemented by tests for further parameters and, when applying the proposed multiplicity adjustment, the adjusted overall power to find some difference will typically remain at a very similar level as the power of the best included test.

\section{Data example}\label{sec.example}
As an illustrating example we considered comparisons from the study ``Pembrolizumab alone or with chemotherapy versus cetuximab with chemotherapy for recurrent or metastatic squamous cell carcinoma of the head and neck (KEYNOTE-048): a randomised, open-label, phase 3 study'' by Burtness et al.\cite{burtness2019pembrolizumab}. The survival curves shown in this study exhibit obvious properties of non-proportional hazards. For the comparison of pembrolizumab alone versus cetuximab with chemotherapy, survival curves were crossing, with better survival under cetuximab with chemotherapy in the first eight months and subsequent better survival under pembrolizumab. In the comparison of pembrolizumab with chemotherapy versus cetuximab with chemotherapy, survival curves were almost equal in both groups for the first eight months with subsequent separation of the two curves, showing a benefit for the pembrolizumab group.

We reconstructed a data set for the comparison of overall survival in the full population under pembrolizumab alone versus
cetuximab with chemotherapy based on the numbers at risk and number of censored event-times, which are given for every five-month interval in Figure 2D of this publication \cite{burtness2019pembrolizumab}. In the reconstructed data set, event times and censoring times, respectively, were equally spread within each five-month interval. The estimated survival functions for the reconstructed data are shown in Figure \ref{fig.bsp_pembrolizumab_2D}. The data set contains 301 subjects with 237 events in the pembrolizumab alone group and 300 subjects with 264 events in the cetuximab with chemotherapy group.
The overall median and maximum follow up times are 0.96 years and 3.96 years.

In the study by Burtness et al.\cite{burtness2019pembrolizumab}, confirmatory tests were performed for the Cox model hazard ratio between groups. As discussed, under non-proportional hazards, the expected value of the Cox model hazard ratio depends on trial characteristics such as the length of recruitment and follow-up periods and other parameters may be more appropriate to quantify the treatment effect, in particular under crossing survival curves (see, e.g. Magirr and Burman\cite{magirr2019modestly}). Accordingly, survival functions were further characterised in Burtness et al. by reporting 0.5-year survival, 1-year survival and median survival times, however no inference for the between-group difference of these parameters and, consequently, no adjustment for simultaneous inference was provided. In our example, assume there is some information from previous studies that the survival curves will likely show late separation and a standard analysis may hence be affected by non-proportional hazards.

We consider two alternative analysis approaches. The first approach is focused on establishing a difference between survival curves based on a small set of parameters, the second one aims at a characterisation of differences by confidence intervals for a larger set of parameters. 

In the first approach, two primary null hypotheses for the overall differences in survival functions and for the difference in two-year survival probabilities are defined and tested, respectively, with the Cox model score test and the Wald test for survival differences. Multiplicity adjustment at the simultaneous one-sided significance level of 0.025 is applied, using the closed-test based on the multivariate normal distribution with asymptotic covariance estimate. This analysis is intended to show at least some benefit of pembrolizumab (group 1) over cetuximab (group 0). Two tests are combined to complement for a possible lack in power of the score test compared to the potentially large difference in survival at a late milestone time-point. The according null hypotheses are $H_{01}: \lambda_1(s)\geq\lambda_0(s), \forall s \leq 0 \leq 3.5$ and $H_{02}: S_1(2) \leq S_0(2)$.

The resulting unadjusted one-sided p-values for $H_{01}$ and $H_{02}$ are $p_1=0.0100$ and $p_2=0.0053$. The according multiplicity adjusted one-sided p-values are $p_{1,adj}=0.0100$ and $p_{2,adj}=0.0082$. Hence, both null hypothesis are rejected at the one-sided family-wise significance level of 0.025. Furthermore, the adjustment did not change the p-value of the score test, and only by a small amount increased the p-value of the test for 2-year survival differences. In this example, the estimated correlation between the two test statistics was 0.87, and this large correlation entails the very modest adjustment to the p-values.

In the second example analysis approach, the difference in survival curves is characterised by a parameter set that includes the differences in 0.5-year survival, 1-year survival, 2-year survival, median survival times and 3.5-year RMST. Simultaneous two-sided 95\% confidence intervals for these parameters are calculated using the multivariate normal adjustment with the asymptotic covariance estimate.

The results of the second analysis are shown in Table \ref{tab.bsp_pembrolizumab_2D}. In this example, the width of Bonferroni-adjusted confidence intervals is 1.064 times the width of multivariate normal adjusted intervals. 

The analysis shows that regarding 0.5-year survival, lower efficacy of pembrolizumab alone or with chemotherapy versus cetuximab cannot be ruled out, with a difference up to 13 percenentage points at the adjusted 95\% confidence level. At 1-year, the relation has reversed with point estimates for survival probabalities and for the median, which is close to one year in both groups, favoring pembrolizumab. Albeit untercertainty remains at this time-point, reflected in the confidence intervals that cover the possibility of no between-groups difference in 1-year survival and in the median survival times. Only at longer time spans point estimates and confidence intervals for 2-year survival and the 3.5-year RMST difference support the conclusion of larger benefit under pembrolizumab. 

\section{Software implementation}
The proposed methods were implemented in the R \cite{R2021} function nphparams(). This function was added to the previously published R library nph \cite{ristl2021delayed}, which provides functions to simulate and analyse survival data under non-proportional hazards. 

The example data set of Section \ref{sec.example}, too, was added to the nph package under the name pembro. The following R code may be applied to reproduce the examplary data analysis:

\begin{verbatim}
#Install and load nph library
  install.packages("nph")
  library(nph)

#Load example data set
    data(pembro)

#Primary hypothesis tests with closed-testing 
#multiplicity adjustment for parameter set 1
    set1<-nphparams(time=time, event=event, group=group,data=pembro,
        param_type=c("score","S"),
        param_par=c(3.5,2),
        param_alternative=c("less","greater"),
        closed_test=TRUE,alternative_test="one.sided")
    print(set1)

#Quantification of differences via simultaneous 95\% confidence intervals
#for parameter set 2 
    set2<-nphparams(time=time, event=event, group=group, data=pembro,  
        param_type=c("S","S","S","Q","RMST"),
        param_par=c(0.5,1,2,0.5,3.5))
        print(set2)

#Recreate Figure 4 (survival curves and indication of chosen parameters)
    plot(set2,trt_name="Pembrolizumab",ctr_name="Cetuximab",
        showlines=TRUE)

    
\end{verbatim}

\clearpage
\section{Discussion}

In absence of the proportional hazard assumption, the exact definition of a survival benefit under treatment versus control is ambiguous. Essentially, two distribution functions need to be compared and different aspects of these distributions may receive different emphasis depending on personal preferences or circumstances. E.g., a survey by Shafrin et al. \cite{shafrin2017patient} among melanoma patients and lung cancer patients and their treating physicians found that patients on average preferred a larger chance for increased long-term survival and in exchange were more willing to accept increased short-time risk for mortality as opposed to their physicians. 

Consequently, to formally establish a benefit of treatment over control in a clinical trial under non-proportional hazards, more than one parameter for the difference in survival functions needs to be regarded. Our aim was to provide a formal inference framework that includes a wide range of suitable parameters and that allows for an efficient parametric multiplicity adjustment to control the type I error rate of hypothesis tests and the simultaneous coverage of confidence intervals.

All considered parameter estimates essentially are a function of the observed event process and as such can be combined in a joint counting process framework that establishes their asymptotic multivariate normal distribution. Simultaneous inference based on this distributional approximation results in more powerful procedures than adjustments such as the Bonferroni-Holm method, which do not take into account the underlying distribution. In particular when parameters are highly correlated, as is, e.g., the case for combinations such as 3-year survival and RMST in our simulation scenarios, only moderate multiplicity adjustment is required and the proposed methods result in little loss in efficacy compared to unadjusted univariate analyses.

Of note, in absence of proportional hazards, the Cox model hazard ratio is not robust to design characteristics as it depends on the timing of events and hence can be affected, e.g., by the recruitment rate and length of follow-up. Thus the traditional hazard ratio is of limited use to quantify differences in survival curves under non-proportional hazards. In particular with crossing survival curves (or more generally crossing hazard curves) the hazard ratio estimate should be interpreted with care. 
The logrank test, or equivalently the Cox model score test, however, is calculated under a global null hypothesis of equal hazard functions and therefore is a viable approach to establish that there is at least some difference between two survival functions.

As an alternative to the proposed multidimensional parametric approach, conclusions could also be drawn from overall inspection of the observed survival functions, and simultaneous inference could be based on confidence bands with simultaneous coverage \cite{parzen1997simultaneous,sachs2022confidence}. This would correspond to a fully non-parametric approach. Such an approach may be suitable to inform the treatment decision of an individual patient, however, it is not suitable to define success criteria regarding efficacy in a clinical trial or when evaluating treatment strategies in clinical practice. To interpret and communicate effects of drugs with a complex pattern of efficacy over time, a set of quantifying parameters seems to be a good compromise between a single parameter, such as the hazard ratio, and a completely non-parametric approach of regarding the overall survival curves.

The multiple testing procedures described in Section \ref{sec:inference} may be extended towards more complex methods. 
A serial gate-keeping procedure \cite{dmitrienko2007gatekeeping} may be used to first show a difference between treatment and control by testing an intersection hypothesis for a small set of parameters for which a high power is expected, and in case of success assess the survival differences in detail with respect to a larger set of relevant parameters. E.g., in the example of Section \ref{sec.example}, the intersection hypothesis test comprising the logrank test and the test for 2-year survival could be used to establish some difference and act as gate-keeper for the subsequent analysis of the larger parameter set. Of note, the rejection of the gate-keeping intersection hypothesis in the first step does not automatically imply the rejection of the corresponding elementary null hypotheses of the first set, because these are not included in the closed testing procedure that corresponds to the gate-keeping approach.
To define more complex testing procedures with several levels, parametric graphical multiple testing procedures \cite{bretz2011graphical,maurer2013memory} could be defined using the estimated covariance matrix.

The first step in our example is similar in spirit to the combined test proposed by Royston and Parmar\cite{royston2016augmenting}. They suggested to perform a maximum combination test that includes the logrank test and RMST differences at several time-points. Royston and Parmar use permutation (or an approximation thereof) to calculate p-values. However, their test can as well be performed within the asymptotic multivariate normal framework we presented, and may there be supplemented by simultaneous confidence intervals for the included RMST differences.

Though not covered in the present work, the simultaneous inference framework may be extended to include stratified analyses. One way to do so, would be to estimate the parameters of interest and their covariance matrix for each stratum separately and then calculate a weighted average of the per-stratum estimates and the corresponding covariance matrix. Weights could correspond to stratum size, however further investigations into the ideal choice of weights may be warranted.

Also, weighted logrank statistics may be included in the inference framework as further extension. One could also consider to perform interim analyses to allow for early stopping \cite{ghosh2022robust, magirr2022design} and adaptations such as sample size reassessment \cite{magirr2016sample} and modification of the set of parameters, e.g., adding milestone analyses at later time points. Depending on which type of data are considered for the adaptations \cite{bauer2004modification, magirr2016sample}, appropriate adaptive tests have to be implemented. 

In summary, simultaneous inference for a predefined set of survival parameters allows for a robust assessment of treatment efficacy under non-proportional hazards. The required multiplicity adjustments can be performed efficiently based on their asymptotic joint normal distribution.

\bibliographystyle{unsrt}
\bibliography{survival}

\clearpage

\begin{table}[htbp]
  \centering
  \caption{Summary of the expressions required for the variance-covariance estimation according to equation (\ref{eq:covariance}) for the considered parameters. The formal definition of the restricted mean survival time estimate $\hat\mu_i$ and the Cox model log hazard ratio estimate $\hat\beta$ are given in Section \ref{sec.estimates}. In the least column, $L$ is the time-point up to which the (average) hazard ratio, RMST difference or logrank test are calculated.}

\begin{tabular}{lllll}
\hline
Parameter of interest $\theta_k$& Per group estimate & $\hat{a}_{k,i}$ & $\hat{H}_{k,i}(s)$ & $t_k$ \\
\hline
Survival difference & $\hat{S}_i(t)$    & $-\hat{S}_i(t)$      &  1     & $t$ \\
Survival ratio & $\log\hat{S}_i(t)$ & $-1$       &    1   & $t$ \\
Cumulative-hazard ratio & $\mathrm{cloglog}\ \hat{S}_i(t)$ & $-1/\log\hat{S}_i(t)$       &    1   & $t$ \\

Quantile difference & $\hat{q}_i(\gamma)$     &    $\frac{1}{-\hat\lambda(\hat{q}_i(\gamma))}$   &   1    & $\hat{q}_i(\gamma)$ \\
Quantile ratio & $\log\hat{q}_i(\gamma)$     &    $\frac{1}{-\hat{q}_i(\gamma)\hat\lambda(\hat{q}_i(\gamma))}$   &  1    & $\hat{q}_i(\gamma)$ \\
Average hazard ratio & $\log\int\limits_0^L\hat{S}_0^{-}(s)\hat{S}_1^{-}(s)d\hat\Lambda(s)$ & $\frac{1}{\log\int_0^L\hat{S}_0^{-}(s)\hat{S}_1^{-}(s)d\hat\Lambda(s)}$   & $\hat{S}_0^{-}(s)\hat{S}_1^{-}(s)$       & $L$ \\
RMST difference & $\hat\mu_i$ &    1   &$ \sum\limits_{t_j\geq s} \Delta_{t_j}\hat{S}_i(t_j) $     & $L$  \\
Logrank test & $\int\limits_0^L\frac{Y_0(s)Y_1(s)}{Y_0(s)+Y_1(s)} \frac{1}{Y_i(s)}dM_i(s) $&
           1 & $\frac{Y_0(s)Y_1(s)}{Y_0(s)+Y_1(s)}$ & $L$\\
Cox model log hazard ratio & $i\hat\beta$ & $\frac{1}{\sum\limits_{i=0}^1 \int\limits_0^L \frac{Y_0(s)Y_1(s)\exp(\hat\beta)}{Y_0(s)+Y_1(s)\exp(\hat\beta)}dN_i(s)}$ &  $\frac{Y_0(s)Y_1(s)\exp(i\hat\beta)}{Y_0(s)+Y_1(s)\exp(\hat\beta)}$ & L\\

\hline
\end{tabular}
  \label{tab.covar}
\end{table}

\clearpage

\begin{table}[htbp]
  \centering
  \caption{Parameter sets in the simulation study. All parameters considered in the simulation are listed and their true values are shown for the scenarios 1 to 6 of Section \ref{sec.scen}. For log-scaled parameters of Section \ref{sec.estimates}, back-transformed values, i.e. ratios, are shown in the Table. The expected value of the Cox model hazard ratio estimate is shown for comparison. For the score test, the expected contribution of an individual to the summed score statistic is shown.
  Cross-marks (x) indicate which parameter is included in which set. }
    \begin{tabular}{cccccccccccccccc}
    \hline
    Parameter & Interpretation & Time   & \multicolumn{6}{c}{Scenario}                  & \multicolumn{7}{c}{Parameter set} \\
    \hline
     &  &   & 1     & 2     & 3     & 4     & 5     & 6     & 1     & 2     & 3     & 4     & 5     & 6     & 7 \\
    \hline
    S     & Survival diff.& 1     & 0.00  & -0.06 & -0.06 & 0.12  & 0.12  & 0.17  &       & x     &       &       & x     &       &  \\
    \hline
    S     &       & 2     & 0.16  & 0.16  & 0.16  & 0.15  & 0.10  & 0.15  &       & x     &       &       & x     &       &  \\
    S     &       & 3     & 0.20  & 0.29  & 0.29  & 0.15  & 0.09  & 0.10  &       & x     &       &       & x     &       &  \\
    logS  & Survival ratio & 1     & 1.00  & 0.92  & 0.92  & 1.19  & 1.34  & 1.35  &       &       & x     &       &       &       &  \\
    logS  &       & 2     & 1.41  & 1.43  & 1.43  & 1.42  & 1.63  & 1.70  &       &       & x     &       &       &       &  \\
    logS  &       & 3     & 2.28  & 3.29  & 3.29  & 1.69  & 1.74  & 2.05  &       &       & x     &       &       &       &  \\
    Q     & 25\% qu. diff. &       & 0.10  & -0.26 & -0.26 & 0.31  & 0.09  & 0.29  &       & x     &       &       &       &       &  \\
    logQ  & 25\% qu. ratio &       & 1.09  & 0.74  & 0.74  & 1.54  & 1.29  & 1.71  &       &       & x     &       &       &       &  \\
    RMST  & RMST diff. & 3     & 0.26  & 0.20  & 0.20  & 0.36  & 0.28  & 0.40  & x     & x     & x     &       &       &       & x \\
    cloglogS & Cumulative HR & 1     & 1.00  & 1.28  & 1.28  & 0.65  & 0.73  & 0.60  &       &       &       & x     &       & x     &  \\
    cloglogS &       & 2     & 0.63  & 0.64  & 0.64  & 0.65  & 0.73  & 0.65  &       &       &       & x     &       & x     &  \\
    cloglogS &       & 3     & 0.56  & 0.43  & 0.43  & 0.65  & 0.74  & 0.69  &       &       &       & x     &       & x     &  \\
   avgHR & Average HR & 3     & 0.67  & 0.75  & 0.75  & 0.65  & 0.74  & 0.62  & x     &       &       & x     &       &       & x \\
    HR    & Cox model HR & 3     & 0.63  & 0.62  & 0.69  & 0.65  & 0.73  & 0.64  &       &       &       &       &       &       &  \\
    score & Score test & 3     & -0.07 & -0.07 & -0.05 & -0.06 & -0.06 & -0.08 &       &       &       &       & x     & x     & x \\
       \hline
    \end{tabular}
  \label{tab:parametersets}
\end{table}

\clearpage

\begin{table}[htbp]
  \centering
  \caption{Type I error rate and power observed in 50,000 simulation runs for scenarios 1-3.}
    \begin{tabular}{rrrrrrrrr}
    \hline
    Scenario & Set   & Parameter & T1E unadj & T1E adj & T1E Holm & Pow unadj & Pow adj & Pow Holm \\
    \hline
    1     & 5     & Any   & 7.56  & 2.73  & 2.21  & 97.2  & 92.5  & 91.1 \\
          &       & S 1   & 2.58  & 0.92  & 0.75  & 2.6   & 2.5   & 2.5 \\
          &       & S 2   & 2.54  & 0.92  & 0.76  & 84.7  & 76.6  & 75.4 \\
          &       & S 3   & 2.59  & 0.95  & 0.79  & 92.3  & 86.2  & 85.2 \\
          &       & Score test & 2.44  & 0.80   & 0.64  & 92.9  & 86.3  & 84.4 \\
          & 6     & Any   & 7.13  & 2.38  & 1.87  & 97.1  & 92.2  & 90.5 \\
          &       & cloglogS 1 & 2.42  & 0.78  & 0.60   & 2.4   & 2.3   & 2.3 \\
          &       & cloglogS 2 & 2.48  & 0.86  & 0.71  & 84.4  & 75.9  & 74.6 \\
          &       & cloglogS 3 & 2.31  & 0.74  & 0.60   & 92.1  & 85.5  & 84.4 \\
          &       & Score test & 2.44  & 0.79  & 0.62  & 92.9  & 86.2  & 84.2 \\
          & 7     & Any   & 3.22  & 2.46  & 1.01  & 92.9  & 91.1  & 84.7 \\
          &       & avgHR & 2.54  & 2.10   & 0.96  & 84.8  & 83.8  & 78.1 \\
          &       & RMST  & 2.54  & 2.10   & 0.95  & 83    & 82.4  & 77.8 \\
          &       & Score test & 2.44  & 2.00   & 0.96  & 92.9  & 91.0   & 84.7 \\
          \hline
    2     & 5     & Any   & 7.41  & 2.65  & 2.13  & 100.0   & 99.9  & 99.9 \\
          &       & S 1   & 2.6   & 0.89  & 0.71  & 0.1   & 0.1   & 0.1 \\
          &       & S 2   & 2.58  & 0.91  & 0.75  & 86.0    & 79.6  & 78.2 \\
          &       & S 3   & 2.51  & 0.87  & 0.73  & 100.0   & 99.9  & 99.9 \\
          &       & Score test & 2.46  & 0.90   & 0.71  & 95.3  & 91.3  & 89.6 \\
          & 6     & Any   & 7.09  & 2.39  & 1.90   & 100.0   & 99.9  & 99.9 \\
          &       & cloglogS 1 & 2.47  & 0.79  & 0.61  & 0.1   & 0.1   & 0.1 \\
          &       & cloglogS 2 & 2.5   & 0.86  & 0.70  & 85.7  & 79.1  & 77.7 \\
          &       & cloglogS 3 & 2.31  & 0.71  & 0.60  & 100.0   & 99.9  & 99.9 \\
          &       & Score test & 2.46  & 0.89  & 0.71  & 95.3  & 91.3  & 89.6 \\
          & 7     & Any   & 3.38  & 2.52  & 1.14  & 95.3  & 93.8  & 89.0 \\
          &       & avgHR & 2.63  & 2.11  & 1.04  & 60.6  & 58.1  & 49.6 \\
          &       & RMST  & 2.62  & 2.11  & 1.04  & 47.9  & 47.6  & 45.1 \\
          &       & Score test & 2.46  & 2.01  & 1.10   & 95.3  & 93.8  & 89.0 \\
          \hline
    3     & 5     & Any   & 7.67  & 2.72  & 2.24  & 99.6  & 98.6  & 98.2 \\
          &       & S 1   & 2.52  & 0.84  & 0.73  & 0.1   & 0.1   & 0.1 \\
          &       & S 2   & 2.45  & 0.86  & 0.70   & 83.6  & 75.2  & 72.3 \\
          &       & S 3   & 2.78  & 1.01  & 0.84  & 99.4  & 98.2  & 97.7 \\
          &       & Score test & 2.31  & 0.87  & 0.70   & 76.7  & 67.6  & 64.7 \\
          & 6     & Any   & 7.06  & 2.27  & 1.80   & 99.6  & 98.4  & 97.9 \\
          &       & cloglogS 1 & 2.41  & 0.77  & 0.63  & 0.1   & 0.1   & 0.1 \\
          &       & cloglogS 2 & 2.39  & 0.78  & 0.62  & 83.2  & 74.6  & 71.6 \\
          &       & cloglogS 3 & 2.21  & 0.62  & 0.49  & 99.3  & 98.0    & 97.4 \\
          &       & Score test & 2.31  & 0.85  & 0.69  & 76.7  & 67.5  & 64.6 \\
          & 7     & Any   & 2.9   & 2.28  & 1.01  & 76.7  & 72.4  & 60.7 \\
          &       & avgHR & 2.43  & 2.05  & 0.98  & 56.9  & 53.8  & 45.7 \\
          &       & RMST  & 2.37  & 2.02  & 0.98  & 45.8  & 45.2  & 42.1 \\
          &       & Score test & 2.31  & 1.96  & 1.00     & 76.7  & 72.4  & 60.7 \\
          \hline

    \end{tabular}
  \label{tab.hyp_test_1}
\end{table}
\clearpage
\begin{table}[htbp]
  \centering
  \caption{Type I error rate and power observed in 50,000 simulation runs for scenarios 4-6.}
    \begin{tabular}{rrrrrrrrr}
    \hline
    Scenario & Set   & Parameter & T1E unadj & T1E adj & T1E Holm & Pow unadj & Pow adj & Pow Holm \\
    \hline
        4     & 5     & Any   & 6.41  & 2.46  & 1.74  & 94.6  & 88.4  & 85.2 \\
          &       & S 1   & 2.39  & 0.94  & 0.73  & 66.8  & 62.6  & 61.3 \\
          &       & S 2   & 2.61  & 1.05  & 0.73  & 83.5  & 77.1  & 74.9 \\
          &       & S 3   & 2.56  & 1.01  & 0.76  & 83.0    & 77.0    & 75.3 \\
          &       & Score test & 2.45  & 0.99  & 0.70   & 91.6  & 85.0  & 81.7 \\
          & 6     & Any   & 6.24  & 2.31  & 1.60   & 94.5  & 88.1  & 84.8 \\
          &       & cloglogS 1 & 2.31  & 0.88  & 0.66  & 66.1  & 61.7  & 60.4 \\
          &       & cloglogS 2 & 2.55  & 0.99  & 0.69  & 83.2  & 76.5  & 74.1 \\
          &       & cloglogS 3 & 2.47  & 0.93  & 0.69  & 82.7  & 76.5  & 74.7 \\
          &       & Score test & 2.45  & 0.98  & 0.69  & 91.6  & 84.9  & 81.6 \\
          & 7     & Any   & 3.27  & 2.40   & 1.12  & 92.8  & 90.9  & 84.7 \\
          &       & avgHR & 2.47  & 1.98  & 1.04  & 90.6  & 89.3  & 83.8 \\
          &       & RMST  & 2.50   & 2.00     & 1.04  & 90.0    & 88.9  & 83.9 \\
          &       & Score test & 2.45  & 1.97  & 1.07  & 91.6  & 90.2  & 84.5 \\
          \hline
    5     & 5     & Any   & 6.30   & 2.58  & 1.79  & 83.9  & 71.7  & 65.9 \\
          &       & S 1   & 2.58  & 1.10  & 0.80   & 63.6  & 54.9  & 51.9 \\
          &       & S 2   & 2.49  & 1.08  & 0.76  & 64.6  & 55.3  & 51.3 \\
          &       & S 3   & 2.60   & 1.11  & 0.79  & 54.0    & 46.9  & 44.2 \\
          &       & Score test & 2.45  & 1.07  & 0.74  & 76.7  & 65.3  & 59.7 \\
          & 6     & Any   & 6.06  & 2.37  & 1.63  & 83.6  & 71.1  & 65.2 \\
          &       & cloglogS 1 & 2.51  & 1.04  & 0.75  & 63.1  & 54.2  & 51.1 \\
          &       & cloglogS 2 & 2.46  & 1.04  & 0.72  & 64.3  & 54.7  & 50.7 \\
          &       & cloglogS 3 & 2.39  & 0.93  & 0.66  & 53.5  & 46.1  & 43.4 \\
          &       & Score test & 2.45  & 1.07  & 0.73  & 76.7  & 65.3  & 59.5 \\
          & 7     & Any   & 3.53  & 2.40   & 1.26  & 79.9  & 75.1  & 65.2 \\
          &       & avgHR & 2.49  & 1.88  & 1.12  & 71.6  & 69.6  & 63.3 \\
          &       & RMST  & 2.49  & 1.88  & 1.03  & 77.0    & 73.4  & 64.1 \\
          &       & Score test & 2.45  & 1.85  & 1.01  & 76.7  & 73.0    & 64.1 \\
          \hline
    6     & 5     & Any   & 7.16  & 2.58  & 1.97  & 98.1  & 94.6  & 93.2 \\
          &       & S 1   & 2.42  & 0.84  & 0.68  & 90.4  & 85.4  & 84.5 \\
          &       & S 2   & 2.63  & 0.95  & 0.73  & 88.0    & 82.2  & 81.1 \\
          &       & S 3   & 2.67  & 1.03  & 0.78  & 56.8  & 54.1  & 53.6 \\
          &       & Score test & 2.51  & 0.93  & 0.73  & 96.7  & 92.7  & 91.0 \\
          & 6     & Any   & 6.58  & 2.17  & 1.65  & 98.1  & 94.4  & 93.0 \\
          &       & cloglogS 1 & 2.35  & 0.77  & 0.64  & 90.1  & 84.8  & 83.8 \\
          &       & cloglogS 2 & 2.57  & 0.89  & 0.68  & 87.7  & 81.7  & 80.6 \\
          &       & cloglogS 3 & 2.11  & 0.65  & 0.48  & 56.4  & 53.4  & 52.9 \\
          &       & Score test & 2.51  & 0.92  & 0.72  & 96.7  & 92.6  & 91.0 \\
          & 7     & Any   & 3.36  & 2.44  & 1.16  & 98.0    & 97.3  & 94.7 \\
          &       & avgHR & 2.52  & 2.03  & 1.10   & 97.5  & 97.0    & 94.6 \\
          &       & RMST  & 2.54  & 2.01  & 1.04  & 97.4  & 96.8  & 94.3 \\
          &       & Score test & 2.51  & 1.97  & 1.05  & 96.7  & 96.3  & 94.3 \\
          \hline
    \end{tabular}
  \label{tab.hyp_test_2}
\end{table}
\clearpage

\begin{table}[htbp]
  \centering
  \caption{Analysis of example data based on Figure 2D from KEYNOTE-048}
    \begin{tabular}{rrrrrrr}
     \hline
        &       &       &     &  \multicolumn{3}{c}{95\% Confidence intervals} \\
      Parameter & Pembro. & Cetux.  & Difference & Undadjusted & MVN adjusted & Bonferroni \\
    \hline
    0.5-year survival & 0.721 & 0.763 & -0.042 & [-0.112, 0.028] & [-0.129, 0.044] & [-0.134, 0.049] \\
    1-year survival & 0.514 & 0.465 & 0.050 & [-0.030, 0.129] & [-0.049, 0.148] & [-0.056, 0.155] \\
    2-year survival & 0.277 & 0.189 & 0.088 & [0.021, 0.155] & [0.005, 0.171] & [-0.001, 0.177] \\
    Median survival & 1.037 & 0.915 & 0.122 & [-0.075, 0.319] & [-0.121, 0.365] & [-0.136, 0.381] \\
    3.5-year RMST & 1.436 & 1.232 & 0.204 & [0.027, 0.381] & [-0.015, 0.422] & [-0.029, 0.437] \\
    \hline
    \end{tabular}
  \label{tab.bsp_pembrolizumab_2D}
\end{table}

\clearpage

\begin{figure}[!p]
    \begin{center}
          \includegraphics[page=2,scale=.6,clip,trim=0cm 29cm 0.00cm 0cm]{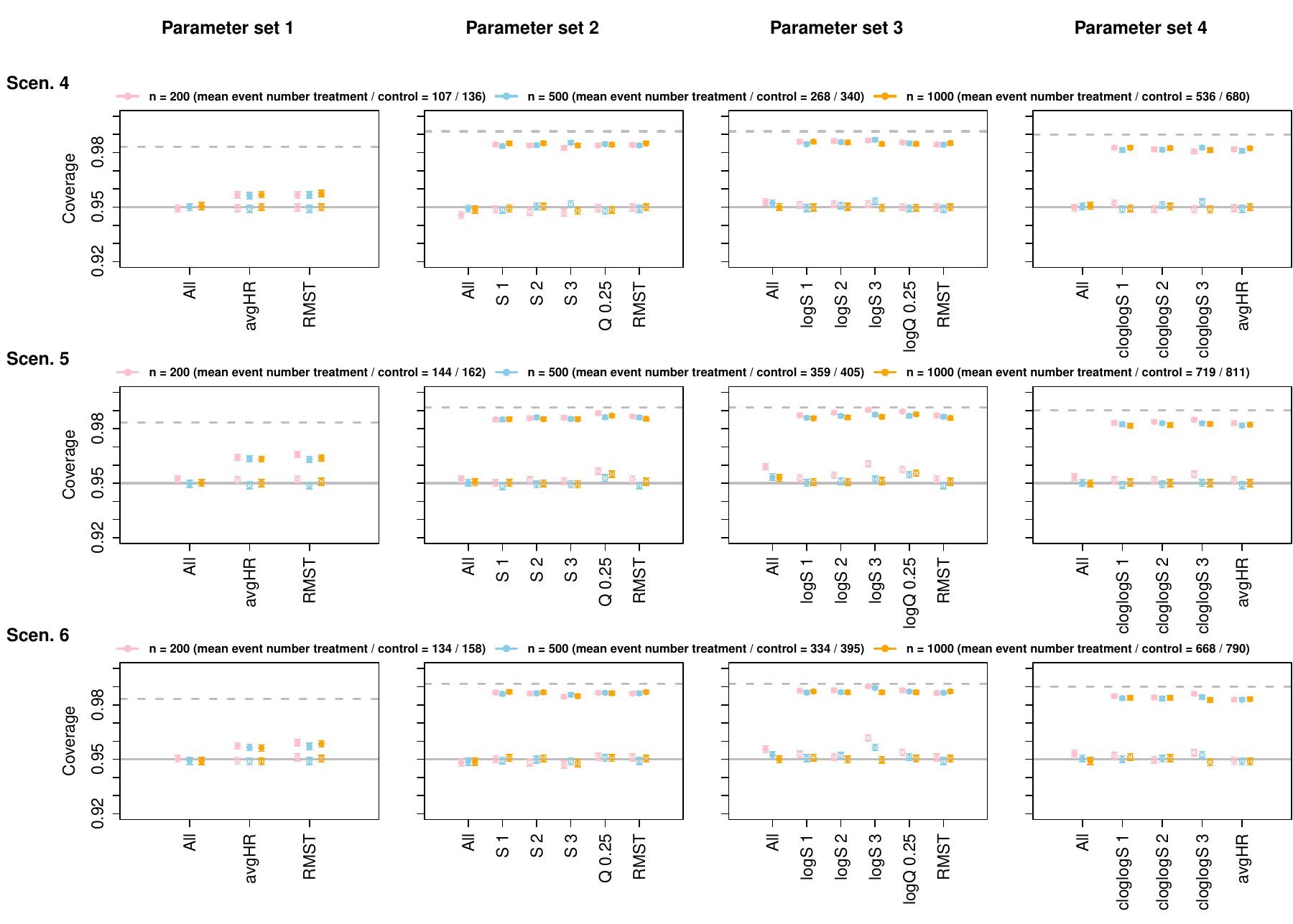} 
    \end{center}
\caption{Empirical coverage of confidence intervals for Scenarios 4-6, based on the multivariate normal distribution with asymptotic covariance matrix estimate. Further details as in Figure \ref{fig.coverage_asym123}.
}
\label{fig.coverage_asym456}
\end{figure}

\begin{figure}[!p]
    \begin{center}
          \includegraphics[page=1,scale=.55,clip,trim=0cm 0cm 0.00cm 0cm]{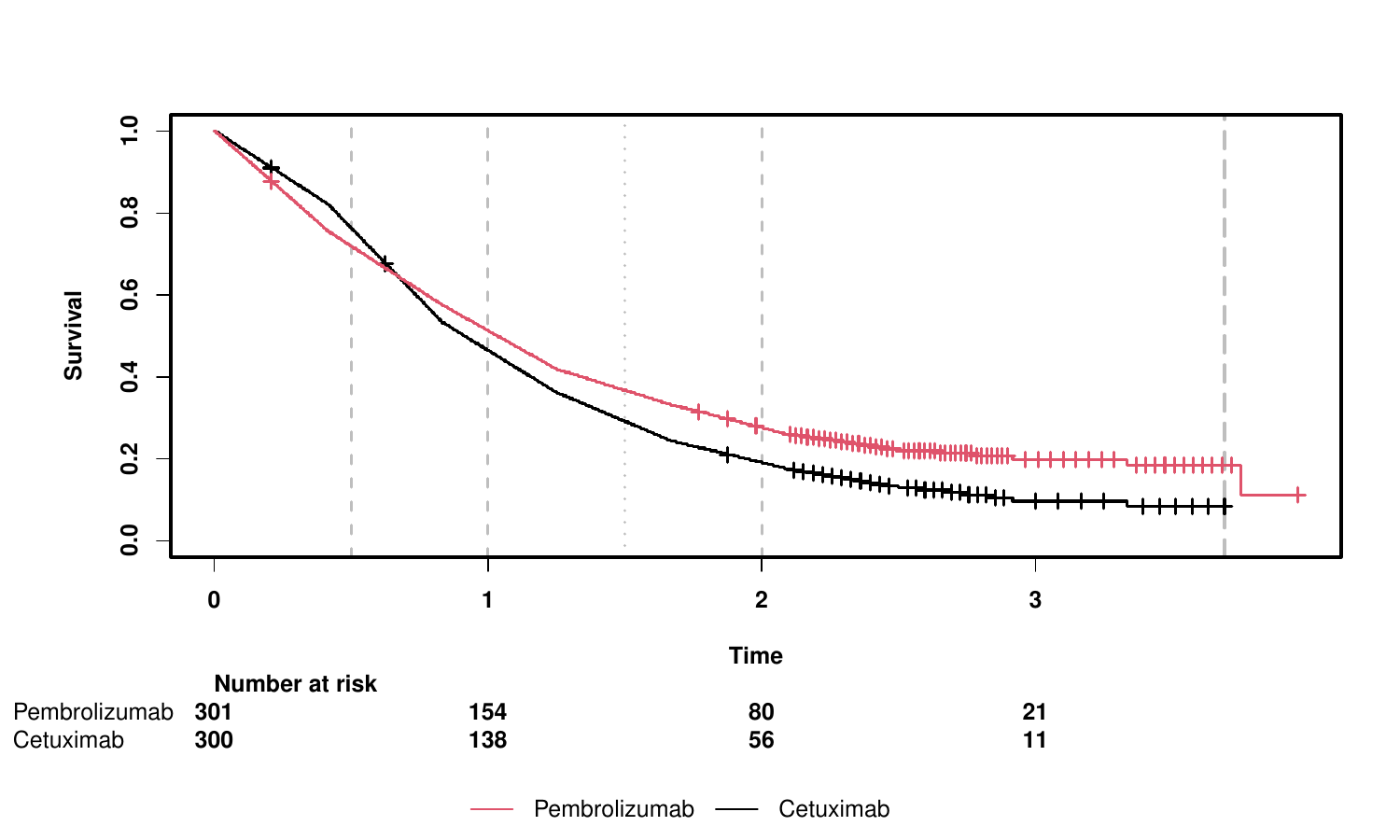}
    \end{center}
\caption{Analysis of example data based on Figure 2D of Burtness et al.  \cite{burtness2019pembrolizumab}.
}
\label{fig.bsp_pembrolizumab_2D}
\end{figure}

\clearpage
\section*{Supplemental material - Simulation results for the perturbation approach}
\renewcommand{\thefigure}{S\arabic{figure}}
\setcounter{figure}{0}
\renewcommand{\thetable}{S\arabic{table}}
\setcounter{table}{0}

\begin{figure}[ht]
    \begin{center}
          \includegraphics[page=1,scale=.55,clip,trim=0cm 29cm 0.00cm 0cm]{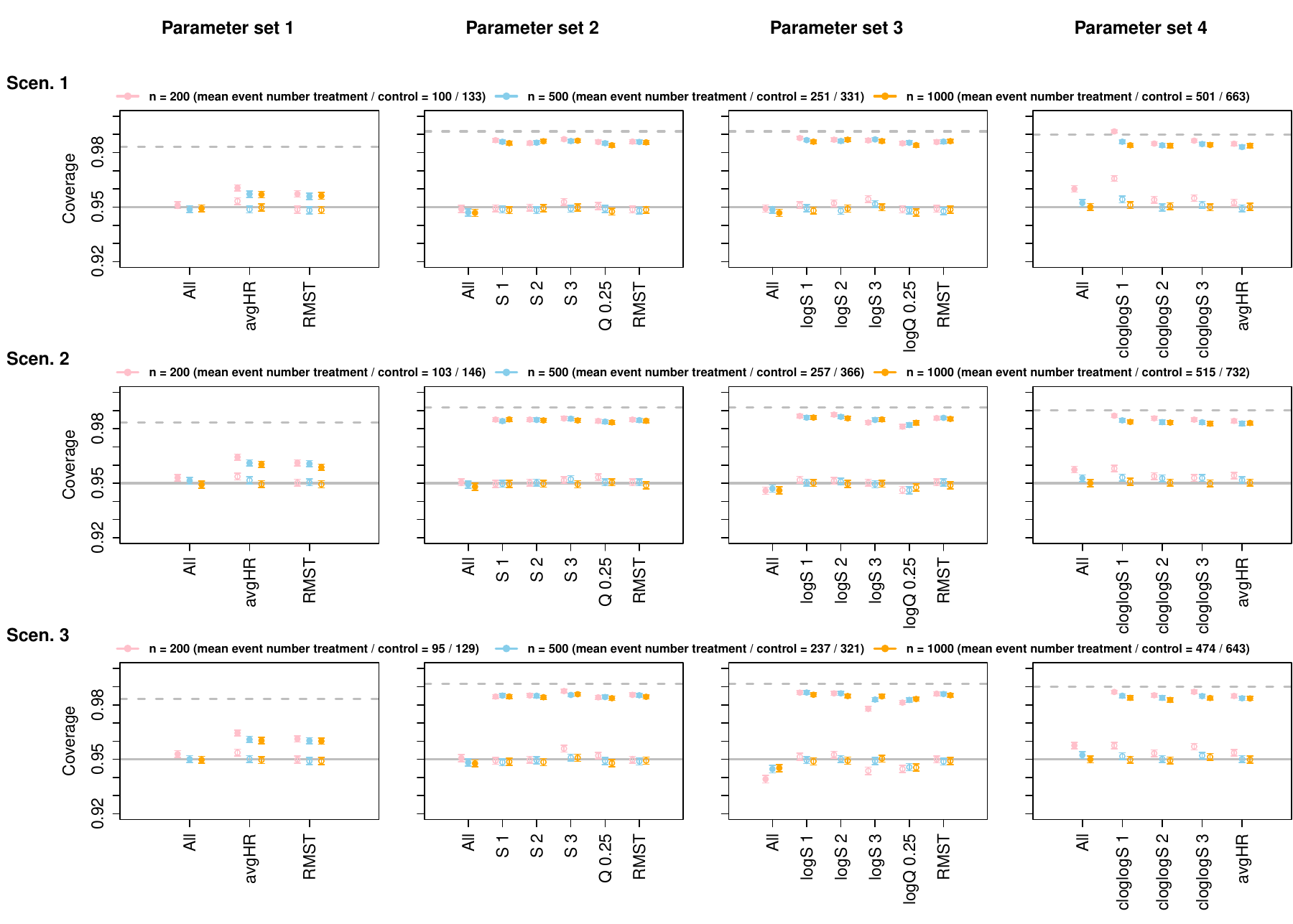}
    \end{center}
\caption{
Empirical coverage of confidence intervals for Scenarios 1-3, based on the multivariate normal distribution with perturbation covariance matrix estimate. Filled circles show the simultaneous coverage (All) and univariate coverage probability of multiplicity adjusted intervals for single parameters (abbreviations as in Table\ref{tab:parametersets}). Open circles represent the coverage of unadjusted univariate confidence intervals. Error bars represent 95\% Wald confidence intervals for the respective coverage probabilities. For comparison, the horizontal solid line indicates the nominal coverage of 95\%, and the horizontal dashed line indicates the univariate confidence level that would result from a Bonferroni adjustment for the respective number of parameters.
}
\label{fig.coverage_perturb123}
\end{figure}

\clearpage
\begin{figure}[ht]
    \begin{center}
          \includegraphics[page=2,scale=.55,clip,trim=0cm 29cm 0.00cm 0cm]{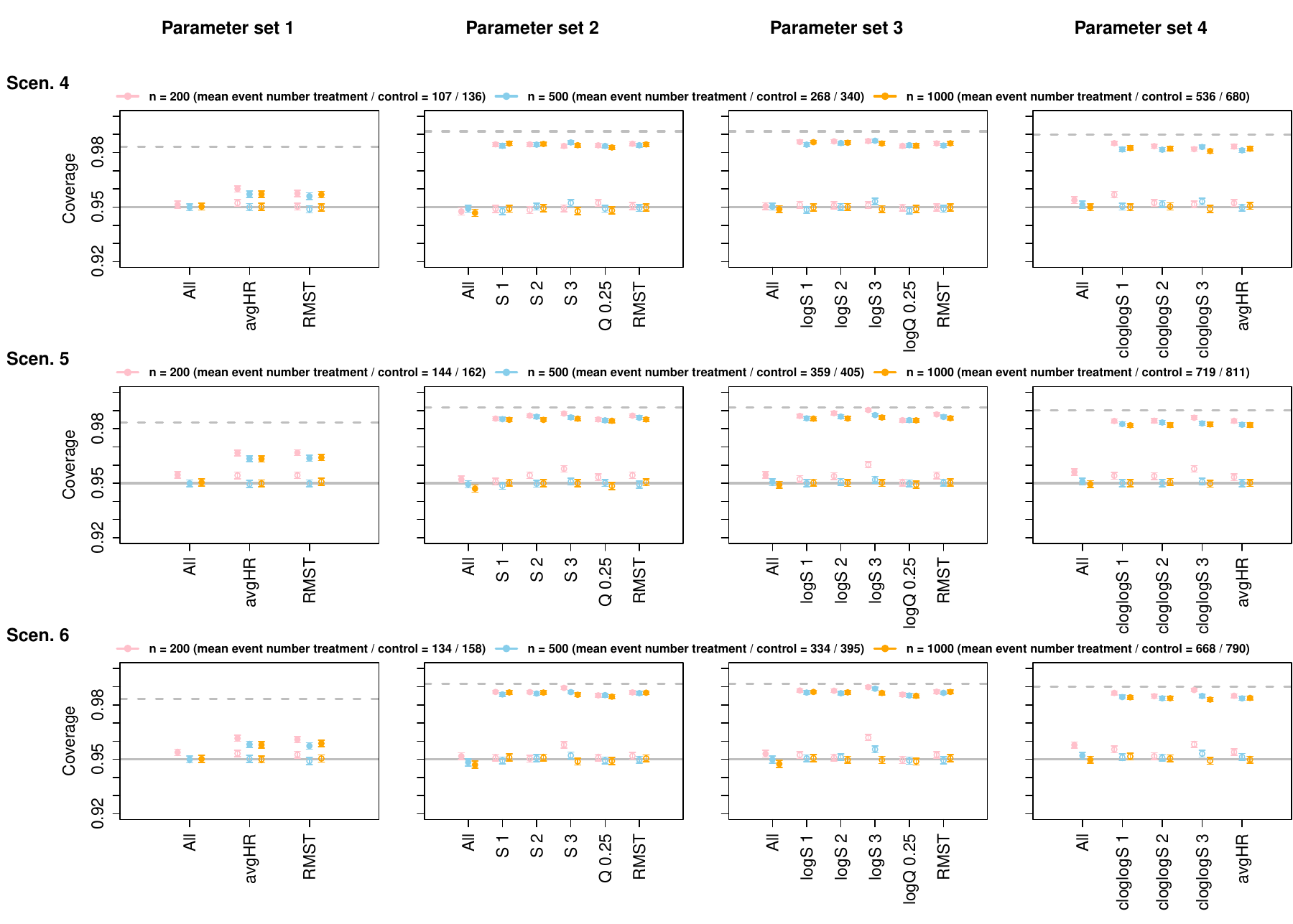}
    \end{center}
\caption{
Empirical coverage of confidence intervals for Scenarios 4-6, based on the multivariate normal distribution with perturbation covariance matrix estimate. Further details as in Figure \ref{fig.coverage_perturb123}.
}
\label{fig.coverage_perturb456}
\end{figure}

\clearpage
\section*{Supplemental material - Simulation results for parameter sets that include the Cox model hazard ratio}

\begin{figure}[ht]
    \begin{center}
          \includegraphics[page=1,scale=.55,clip,trim=0cm 29cm 0.00cm 0cm]{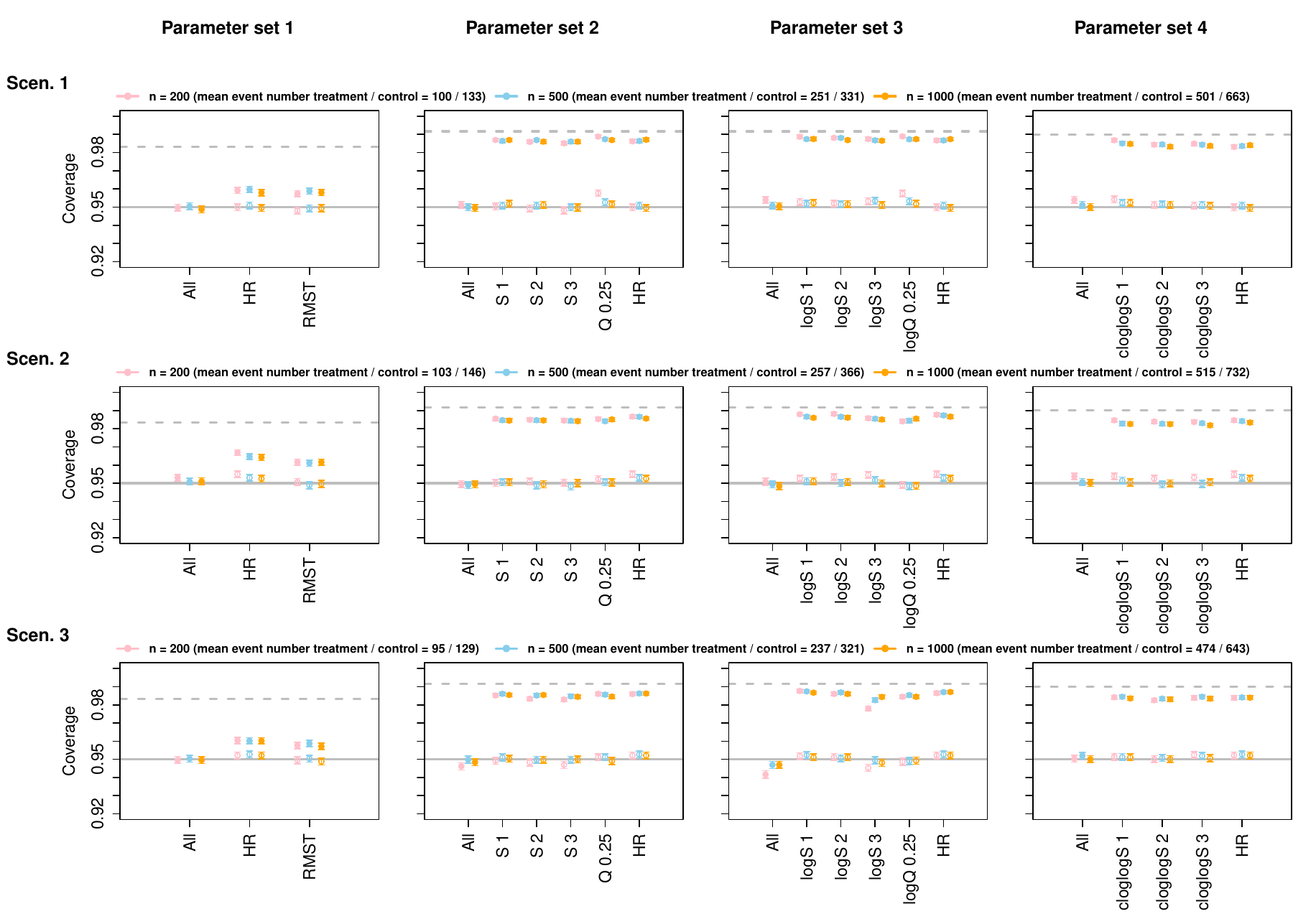}
    \end{center}
\caption{
Empirical coverage of confidence intervals for parameter sets including the Cox model hazard ratio for Scenarios 1-3. Intervals are based on the multivariate normal distribution with asymptotic covariance matrix estimate. In case of the hazard ratio, coverage was calculated for the expected value of the hazard ratio under the specific scenario assumptions. Further details as in Figure \ref{fig.coverage_perturb123}.
}
\label{fig.coverage_asym_HR123}
\end{figure}

\clearpage

\begin{figure}[ht]
    \begin{center}
          \includegraphics[page=2,scale=.55,clip,trim=0cm 29cm 0.00cm 0cm]{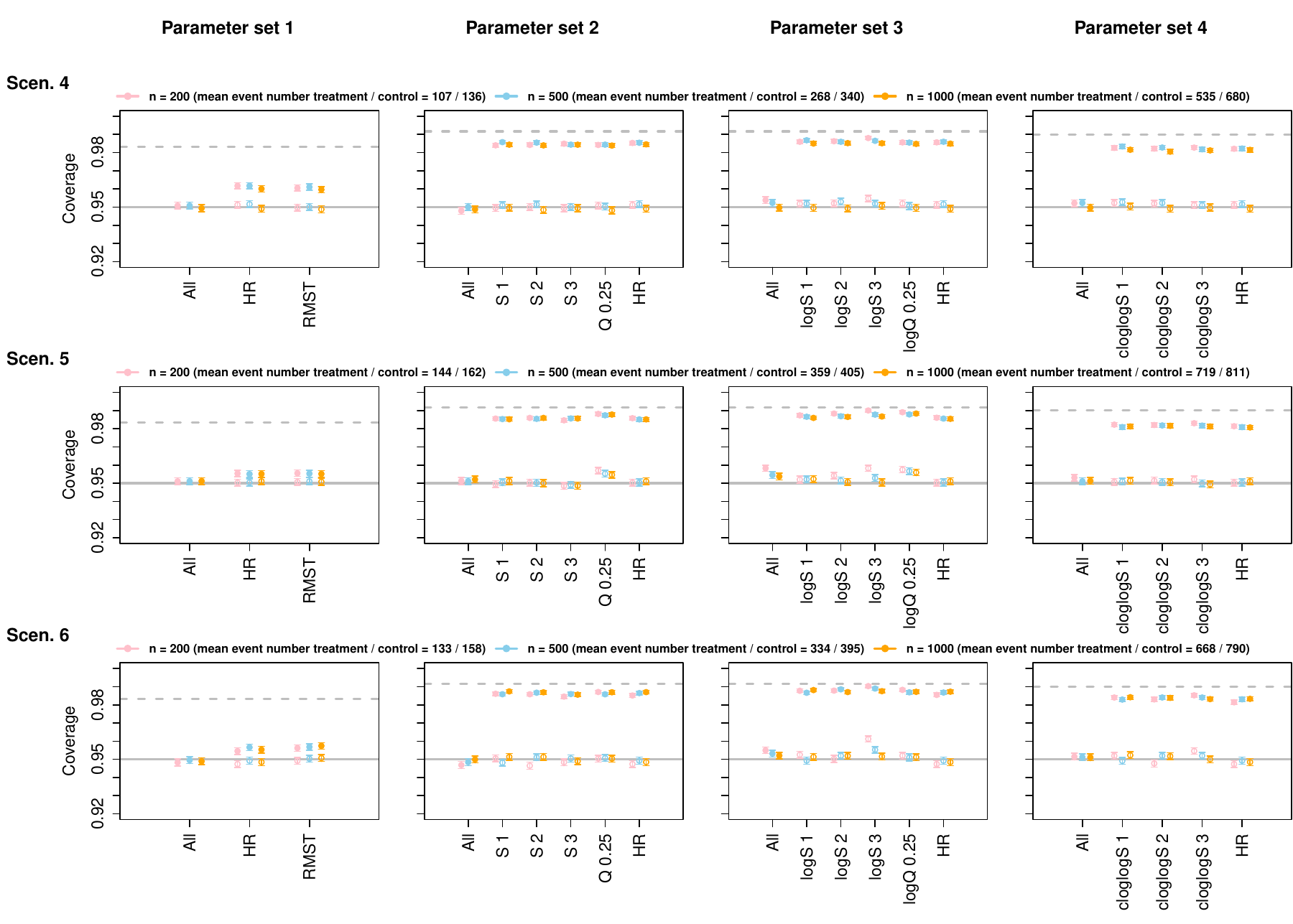}
    \end{center}
\caption{
Empirical coverage of confidence intervals for parameter sets including the Cox model hazard ratio for Scenarios 4-6. Intervals are based on the multivariate normal distribution with asymptotic covariance matrix estimate. In case of the hazard ratio, coverage was calculated for the expected value of the hazard ratio under the specific scenario assumptions. Further details as in Figure \ref{fig.coverage_perturb123}.
}
\label{fig.coverage_asym_HR456}
\end{figure}

\end{document}